\pdfoutput=1

\documentclass[11pt,twoside,a4paper,cmspaper,final,collab]{cms-tdr}
\def\svnVersion{422517P}\def\cmsCernNoTag{CERN-EP-2017-102}\def\cmsCernDate{\today}\def\cmsMessage{Published in the Journal of High Energy Physics as \href{http://dx.doi.org/10.1007/JHEP08(2017)073}{\doi{10.1007/JHEP08(2017)073.}}}

\begin{document}\cmsNoteHeader{B2G-15-006}

\hyphenation{had-ron-i-za-tion}
\hyphenation{cal-or-i-me-ter}
\hyphenation{de-vices}
\RCS$HeadURL: svn+ssh://svn.cern.ch/reps/tdr2/papers/B2G-15-006/trunk/B2G-15-006.tex $
\RCS$Id: B2G-15-006.tex 422456 2017-08-25 15:20:12Z sisagir $
\newlength\cmsFigWidth
\ifthenelse{\boolean{cms@external}}{\setlength\cmsFigWidth{0.85\columnwidth}}{\setlength\cmsFigWidth{0.4\textwidth}}
\ifthenelse{\boolean{cms@external}}{\providecommand{\cmsLeft}{top\xspace}}{\providecommand{\cmsLeft}{left\xspace}}
\ifthenelse{\boolean{cms@external}}{\providecommand{\cmsRight}{bottom\xspace}}{\providecommand{\cmsRight}{right\xspace}}

\newcommand{\xft}{\ensuremath{\mathrm{X}_{5/3}}\xspace}
\newcommand{\dr}{\ensuremath{\Delta_{R}}}
\newcommand{\gt}{$>$}
\newcommand{\HTl}{\ensuremath{H^{\text{lep}}_{\mathrm{T}}}}
\newcommand*{\MADSPIN}{\textsc{MadSpin}\xspace}
\newcommand{\x}{\ensuremath{\phantom{0}}}

\cmsNoteHeader{B2G-15-006}
\title{Search for top quark partners with charge 5/3 in proton-proton collisions at $\sqrt{s}=13\TeV$}

\date{\today}

\abstract{A search for the production of heavy partners of the top quark with charge 5/3 ($\mathrm{X}_{5/3}$) decaying into a top quark and a W boson is performed with a data sample corresponding to an integrated luminosity of 2.3\fbinv, collected in proton-proton collisions at a center-of-mass energy of 13\TeV with the CMS detector at the CERN LHC. Final states with either a pair of same-sign leptons or a single lepton, along with jets, are considered. No significant excess is observed in the data above the expected standard model background contribution and an $\mathrm{X}_{5/3}$ quark with right-handed (left-handed) couplings is excluded at 95\% confidence level for masses below 1020 (990)\GeV. These are the first limits based on a combination of the same-sign dilepton and the single-lepton final states, as well as the most stringent limits on the $\mathrm{X}_{5/3}$ mass to date.}

\hypersetup{%
pdfauthor={CMS Collaboration},%
pdftitle={Search for top quark partners with charge 5/3 in proton-proton collisions at sqrt(s) =13 TeV},%
pdfsubject={CMS},%
pdfkeywords={CMS, BSM, top quark partners, vector-like quarks, hierarchy problem}}

\maketitle

\section{Introduction}
Various extensions of the standard model (SM) predict new heavy particles for addressing 
the hierarchy problem caused by the quadratic divergences in the quantum-loop corrections 
to the Higgs boson (\PH) mass. The largest corrections, owing to the top quark loop, 
are canceled in many of these models, for example composite Higgs models~\cite{TopPartnerHunterGuide, Mrazek:2009yu, Dissertori:2010ug, Contino:2008hi},
by the presence of heavy partners of the top quark. 
This paper describes a search for such spin 1/2 top quark partners,
using data collected by the CMS experiment at $\sqrt{s}=13\TeV$ in 2015.
We focus on a top quark partner with exotic charge +5/3 (in units of the absolute charge of the electron).
Such exotically charged fermions need not necessarily contribute to the coupling of the Higgs boson to gluons~\cite{Azatov:2011qy}, 
and thus the measurements of the Higgs production rates at the LHC set no constraint on the \xft particle. 
While our previous searches and other literature referred to this particle as $\mathrm{T}_{5/3}$, 
in this paper we follow the nomenclature of Ref.~\cite{TopPartnerHunterGuide} and refer to it as $\mathrm{X}_{5/3}$.

\begin{figure}[!htb]
\centering
\includegraphics[width=0.45\textwidth]{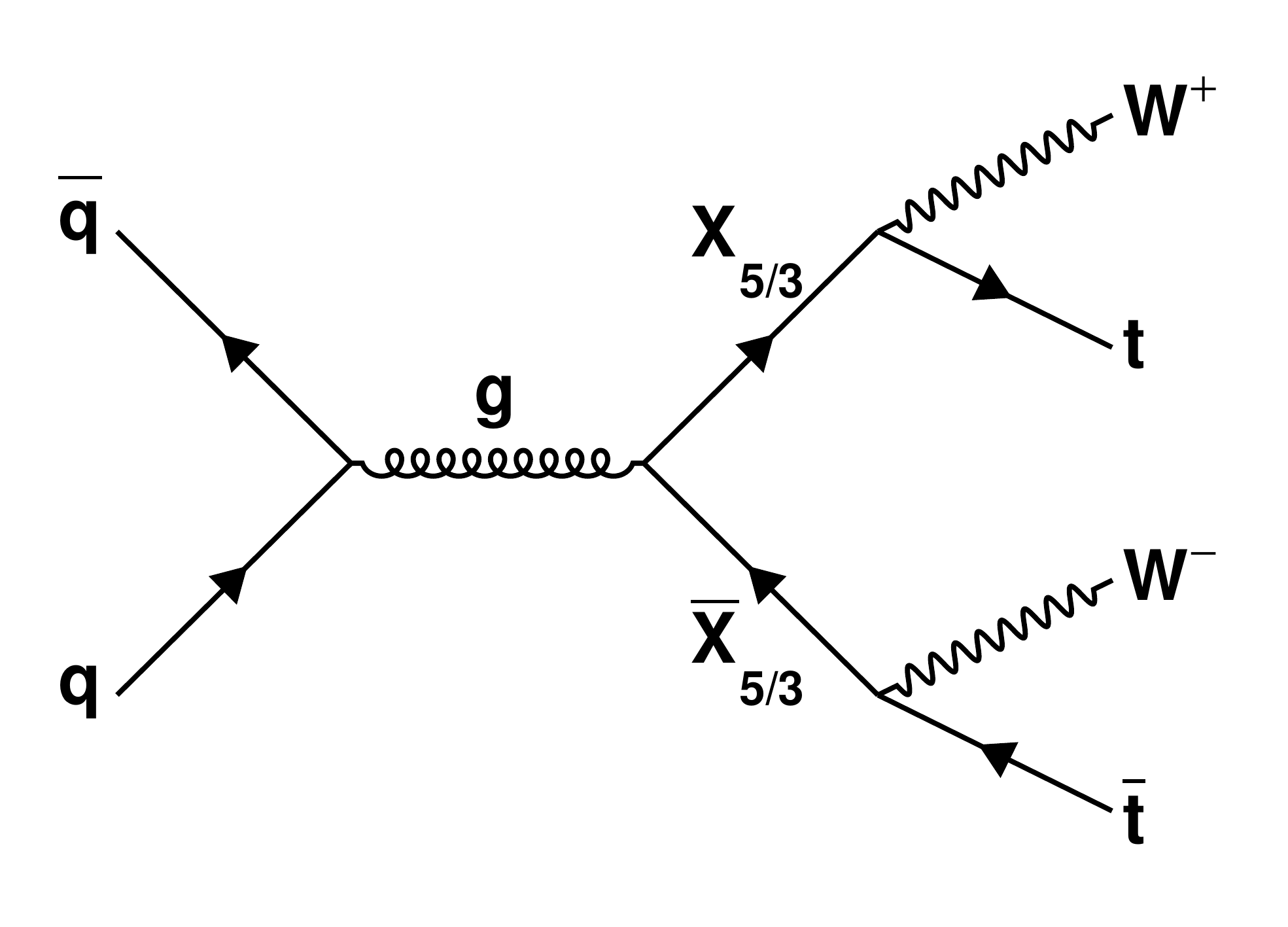}
\includegraphics[width=0.45\textwidth]{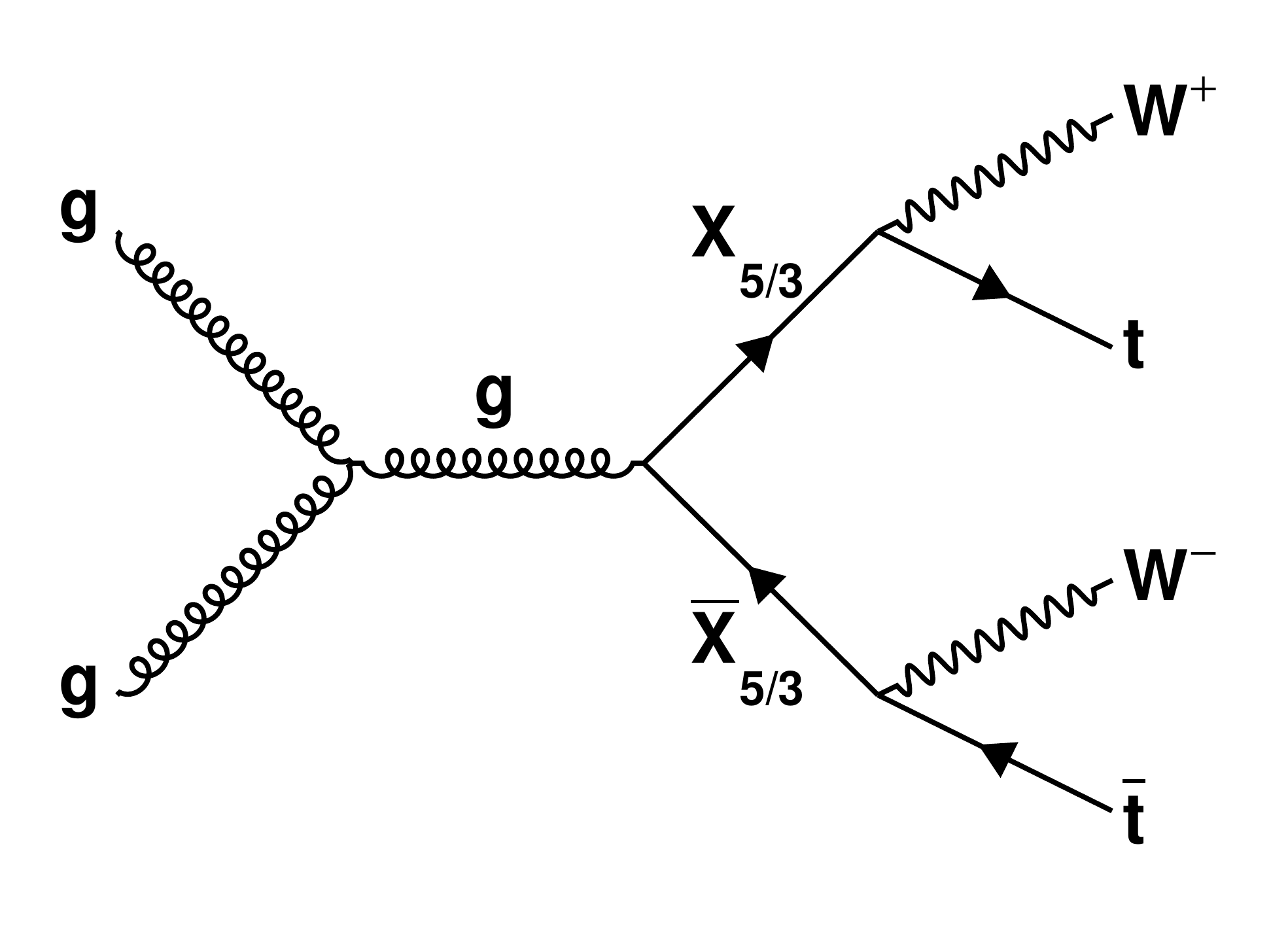}
\caption{Leading order Feynman diagrams for the production and decay of pairs of \xft particles via QCD processes.}
\label{fig:FeynDiag}
\end{figure}

The color charge of the \xft quark allows it to be produced via quantum chromodynamics (QCD) interactions 
in proton-proton collisions with leading-order cross sections that depend on new physics only via the \xft mass. 
We assume that the \xft quark decays via $\xft \rightarrow  \mathrm{t}\PW^+$ followed by $\mathrm{t} \rightarrow \PW^+\mathrm{b}$ 
(charge conjugate modes are implied throughout), which is the dominant decay mode in most models. 
Because mixing of the \xft quark with the top quark only occurs through the weak interaction, production via QCD processes 
always results in the production of \xft pairs (particle and antiparticle), as shown in Fig.~\ref{fig:FeynDiag}. 
The \xft quark can also be produced singly in association with a top quark through electroweak processes; 
however, this production mode is not considered here.

In this paper, the search for the \xft particle is focused on two final states. 
In the ``same-sign dilepton channel'' the two (same-charge) $\PW$ bosons arising from one of the \xft particles 
decay into leptons of the same charge while the other two $\PW$ bosons decay inclusively. 
In the ``single-lepton channel'', one of the $\PW$ bosons decays leptonically into a lepton and a neutrino, 
while the other three $\PW$ bosons decay hadronically (including $\PW\rightarrow\tau\rightarrow$ hadrons).
Throughout the paper, when referring to a lepton ($\ell$), we mean either an electron or muon.
In both channels, leptonic decays from taus are included in the signal region although the lepton identification criteria are optimized for direct decays to either electrons or muons.

A previous search in the same-sign dilepton channel conducted by CMS, using 19.5\fbinv of 
data collected at $\sqrt{s}=8\TeV$, set 
a lower limit on the \xft mass of 800\GeV~\cite{8TeVPaper} at 95\% confidence level (CL). 
Searches have also been performed 
by the ATLAS experiment using 20.3\fbinv of data collected at $\sqrt{s}=8\TeV$ in the same-sign dilepton~\cite{Aad:2015gdg}
and single-lepton~\cite{Aad:2015mba} final states separately, setting lower limits of 740 and 840\GeV, respectively. 

\section{The CMS detector}

The central feature of the CMS apparatus is a superconducting solenoid of 6\unit{m} internal diameter, 
providing a magnetic field of 3.8\unit{T}. Within the solenoid volume are a silicon pixel 
and strip tracker, a lead tungstate crystal electromagnetic calorimeter (ECAL), and a brass and scintillator 
hadron calorimeter (HCAL), each composed of a barrel and two endcap sections. Forward calorimeters extend 
the pseudorapidity ($\eta$) coverage provided by the barrel and endcap detectors. 
Muons are measured in gas-ionization detectors embedded in the steel flux-return yoke outside the solenoid.
The first level of the CMS trigger system, composed of custom hardware processors, 
selects the most interesting events in a fixed time interval of less than 4\unit{\mus}, 
using information from the calorimeters and muon detectors. The high-level trigger processor farm further decreases 
the event rate to a few hundred\unit{Hz}, before data storage. 
A more detailed description of the CMS detector, together with a definition of the coordinate system used 
and the relevant kinematic variables, can be found in Ref.~\cite{myChatrchyan:2008zzk}.

\section{Simulation}

{\tolerance=500
The \xft signal processes are generated using a combination of
\MADGRAPH{}5\_a\MCATNLO 2.2.2~\cite{Alwall:2014hca} and \MADSPIN~\cite{MadSpin}
for two coupling scenarios, corresponding to purely left- or right-handed \xft coupling
to W bosons, denoted by LH and RH, respectively.
The \MADGRAPH generator is used both to produce \xft events and decay each
\xft to a top quark and a W boson, while the decays of the top quarks and W bosons are
simulated with \MADSPIN. The signal events are simulated at leading order (LO) for various mass values 
between 700 and 1600\GeV in 100\GeV steps, separately for each coupling scenario. 
The \xft cross sections are then normalized to the next-to-next-to-leading order 
using Top++2.0~\cite{TPRIMEXSEC,MITOV1,MITOV2,MITOV3,BARNREUTHER,NNLL}.
\par}

The Monte Carlo (MC) background processes are generated with a variety of event generators.
The \MADGRAPH{}5\_a\MCATNLO event generator is used to simulate Z+jets, W+jets, 
single top in the \textit{s}- and \textit{t}-channels, $\ttbar\PZ$, $\ttbar\PW$, $\ttbar\PH$, and $\ttbar\ttbar$ processes, 
as well as events with a combination of three $\PW$ or $\PZ$ bosons and QCD multijet events.
The W+jets and multijet events are generated at LO using the MLM matching scheme~\cite{Alwall:2007fs},
while the others are simulated to next-to-leading order (NLO) using the MLM matching scheme, 
except for Z+jets and $\ttbar\PW$ where the FxFx matching scheme~\cite{Frederix:2012ps}
is used. The \POWHEG 2.0~\cite{Nason:2004rx,Frixione:2007vw,Alioli:2010xd,Alioli:2011as} event generator is used to simulate
$\ttbar$ and single top quark events in the $\PQt\PW$ channel at NLO accuracy.
The diboson events involving $\PW$ or $\PZ$ are generated at LO using either
\MADGRAPH{}5\_a\MCATNLO or \PYTHIA 8.212~\cite{PYTHIA8,PYTHIA82}. 
Parton showering, hadronization, and the underlying event are simulated with 
\PYTHIA, using NNPDF 3.0~\cite{NNPDF30} parton distribution functions (PDF) 
with the CUETP8M1 underlying event tune~\cite{Skands:2014pea}.

All MC events are processed with \GEANTfour~\cite{GEANT,Allison:2006ve} for a full simulation of the CMS detector.
Further, for all simulated samples, additional proton-proton interactions (pileup) are modeled 
by superimposing generated minimum bias interactions onto both the bunch crossing of the simulated events and also in adjacent bunch crossings.
A reweighting procedure is used to match the simulated distributions to the number of pileup interactions observed in data.

\section{Object reconstruction}
\label{sec:Reconstruction}
The analyses described in this paper rely on the reconstruction of four types of objects: 
electrons, muons, jets, and missing transverse energy (\MET). Events are reconstructed using the particle-flow (PF) 
approach~\cite{pfnew}, which consists of reconstructing and identifying each single 
particle with an optimized combination of all subdetector information. 
The details of the object selection are provided below.

Candidate events are required to have at least one reconstructed vertex. 
For events in which there are multiple reconstructed vertices, 
the one with the the largest sum of squared transverse momenta of associated tracks is chosen
as the primary vertex.
For the dilepton analysis, at least two leptons are required to be within the 
tracker acceptance ($|\eta| <2.4$) and to have passed triggers based on dielectron, 
dimuon or electron-muon requirements. 
All double lepton triggers used have an $|\eta| <2.4$ requirement and \pt requirements ranging from 17 to 27\GeV on the leading lepton and from 8 to 12\GeV on the sub-leading lepton. 
The single-lepton analysis requires events to have 
passed a single-electron trigger ($|\eta| <2.1$, $\pt > 27\GeV$) or a single-muon trigger ($|\eta| <2.4$, $ \pt > 20\GeV$).

Electron candidates are reconstructed from a collection of electromagnetic clusters and 
matched to tracks in the tracker~\cite{8TeV-EGamma}. 
They are then required to satisfy identification and isolation criteria. 
The identification criteria make use of shower shape variables, 
track quality requirements, the distance from the track to the primary vertex, and variables measuring compatibility between the track and matched 
electromagnetic clusters to select good electron candidates. Requirements are also imposed to reject 
electrons produced in photon conversions in the detector material.
The isolation variable ($I_{\rm{mini}}$) is defined as the sum of 
energy around the electron in a cone of varying size, divided by the transverse momentum (\pt) of the electron. 
The radius used for the isolation cone ($R$) is defined as:
\[ R = \frac{10\GeV}{\mathrm{min}[\mathrm{max}(\pt,50 \GeV),200\GeV]}\,.\]
We define a ``tight'' (``loose'') electron to have $I_{\rm{mini}} < 0.1\,(0.4)$. 

For the same-sign dilepton analysis, charge misidentification is significantly reduced by 
requiring that different charge measurements for an electron agree (a ${\sim}50\%$ reduction is possible for requiring all measurements agree for low \pt electrons). 
Two of the measurements are based on two different tracking algorithms: the standard CMS track reconstruction 
algorithm~\cite{ctf} and the Gaussian-sum filter algorithm~\cite{gsf}, optimized to take 
into account the possible emission of bremsstrahlung photons in the silicon tracker. 
The third measurement is based on the relative position of the calorimeter cluster and 
the projected track from the pixel detector seed (the pixel hits used to reconstruct an electron's track).  
We find good agreement between the three measurements 
for electrons with $ \pt < 100\GeV$. 
However, for higher-momentum electrons, requiring that the third measurement
agree with the two track-based determinations leads to
a 5--10\% loss in signal efficiency.
Further, the third measurement is also often incorrect for high \pt electrons.
We therefore define a ``relaxed'' charge consistency requirement where for electrons with \pt below 100\GeV 
all three charge measurements are required to agree, while above 100\GeV only the first two measurements 
are required to agree and the third charge measurement is ignored.

Muons are reconstructed using a global track fit of hits in the muon detectors and hits in the silicon tracker. 
The track associated with a muon candidate is required to have at least six hits in the silicon tracker, 
at least one pixel detector hit, and a good quality global fit,
including at least 
one hit in the muon detector. The isolation variable for muons is calculated in the same way as it is for electrons, as described above.
We define a category of ``tight'' muons that satisfy $I_{\rm{mini}} < 0.2$. A second category of ``loose'' muons requires $I_{\rm{mini}} < 0.4$ 
with somewhat relaxed identification requirements.
Additional requirements are imposed on the minimum longitudinal distance of the tracker track with respect 
to the primary vertex ($d_z < 5\unit{mm}$) and the minimum radial distance from the track to the primary 
vertex ($d_{\rm{xy}} < 2\unit{mm}$).

An event-by-event correction using the effective area method~\cite{FastJet1} is applied to the computation of the electron and muon isolation in order to 
account for the effect of pileup. Scale factors to correct for imperfect detector simulation are 
obtained using the ``tag-and-probe'' method~\cite{CMS:2011aa} for lepton identification and isolation, 
as a function of lepton \pt and $\eta$. These scale factors are normally within a few percent of unity and those falling outside that range tend to be consistent with unity.

Jets are clustered from the reconstructed PF candidates using the anti-$k_{t}$ algorithm~\cite{Cacciari:2008gp,FastJet1,FastJet2,FastJet3} with a distance 
parameter of $0.4$ (AK4) and are required to satisfy $\pt > 30\GeV$ and $|\eta| < 2.4$. 
Additional selection criteria are applied to remove spurious features originating from isolated 
noise patterns in certain HCAL regions and from anomalous signals caused by particles depositing 
energy in the silicon avalanche photodiodes used in the ECAL barrel region. 
Jets that overlap with leptons have the leptons removed by matching lepton 
PF candidates to jet constituents and subtracting the energy and momentum of the matched candidates from the jet four-vector. 
Jet energy corrections are applied for residual nonuniformity, nonlinearity of 
the detector response, and the level of pileup in the event~\cite{JEC}.

The missing transverse momentum (\ptvecmiss) is reconstructed as the negative of the vector \pt sum
of all reconstructed PF candidates in an event and its magnitude is denoted as \MET.
Energy scale corrections applied to jets are also propagated to \MET.

\section{Same-sign dilepton final state}

The \xft search in the dilepton channel takes advantage of the same-sign leptons in the final state as well as the significant amount of jet activity due to the presence of the two bottom quarks and the possibility of hadronic decays for one of the top quark partners.

The background contributions associated with this channel fall into three main categories:
\begin{itemize}
\item Same-sign prompt (SSP) leptons: SM processes leading to prompt, same-sign dilepton signatures, where a prompt lepton is defined as one originating from the prompt decay of either a $\PW$ or $\PZ$ boson. Their contribution is obtained from simulation.
\item Opposite-sign prompt leptons: prompt leptons can be misreconstructed with the wrong charge leading to a same-sign dilepton final state. This contribution is estimated using a data-driven method.
\item Same-sign events arising from the presence of one or more non-prompt leptons: this is the primary instrumental background arising from jets misidentified as leptons, 
non-prompt leptons from heavy flavor decays, fake leptons from conversions,  etc. This contribution is also estimated using a data-driven method.
\end{itemize}

After requiring two tight, same-sign leptons with $ \pt > 30\GeV $ we impose the following requirements:

\begin{itemize}
\item Quarkonia veto: require invariant dilepton mass $M_{\ell\ell}$ $> 20\GeV$.
\item Associated $\PZ$ boson veto: ignore any event where $M_{\ell\ell'}$ is within 15\GeV of the mass of the $\PZ$ boson, where $\ell$ is either lepton in the same-sign pair, and $\ell'$ is any lepton not in the same-sign pair, but of the same flavor as the first, and with $ \pt > 30\GeV $.
\item Primary $\PZ$ boson veto: events are rejected if  $76.1 < M_{\ell\ell}<106.1\GeV$ for the dielectron channel only. If the muon charge is mismeasured, its momentum will also be mismeasured, so a selected muon pair from a $\PZ$ boson is unlikely to fall within this invariant mass range.
\item Leading lepton $\pt>40\GeV$.
\item Number of constituents ${\geq}5$.
\item $\HTl >900\GeV$.
\end{itemize}

The ``number of constituents'' is defined as the number of AK4 jets in the event passing our jet selection together with the number of other (i.e. not in the same-sign pair) tight leptons with $ \pt > 30\GeV$. The $\HTl$ used in this analysis is the scalar sum of the \pt of all selected jets and tight leptons in the event. With these requirements we find typical signal efficiencies of roughly $40$ to $50\%$ and background rejection of greater than $99\%$.

\subsection{Background modeling}

\subsubsection{Same-sign prompt lepton background}
The same-sign prompt lepton background consists of contributions from diboson production ($\PW\PZ$ and $\PZ\PZ$) and rarer processes, such as $\ttbar\PW$, $\ttbar \PZ$, $\ttbar\PH$, $\PW\PW\PZ$, $\PZ\PZ\PZ$, $\PW\PZ\PZ$, and $\PW\PW$+jets. Many of these processes have not been observed at the LHC or are not yet well measured. We estimate the contribution from SM events with two prompt same-sign leptons using simulation (see Table~\ref{tab:SummaryYields}).

\subsubsection{Opposite-sign prompt lepton background}
\label{sec:chargemisid}
Processes with two oppositely-charged prompt leptons can contribute to the background if the charge of one of the leptons is incorrectly measured (this background is referred to throughout as ``ChargeMisID''). For muons in the \pt range considered in this analysis, the charge misidentification probability is found to be negligible~\cite{susy2015}. For electrons, the magnitude of this contribution can be derived from data by using a sample dominated by $\PZ$+jets events.
The measurement is performed by first selecting pairs of electrons, with each electron of the pair being in the same $|\eta|$ region and having $\pt < 100\GeV$.  
Each pair is then required to have an invariant mass within 10\GeV of the Z boson mass.
Since the momentum and energy measurements of the electrons are driven by the ECAL information, the pair's invariant mass is insensitive to potential track mismeasurement.
Counting the number of pairs with same-sign charges then provides the charge misidentification probability as a function of $|\eta|$ for electrons with $\pt<100\GeV$.
Next, pairs are formed using one electron with \pt less than 100\GeV and one above 100\GeV.
Again the number of same-sign pairs is counted to determine the charge misidentification probability; making use of the previously measured probability for electrons with $\pt<100\GeV$ then gives a measurement of the charge misidentification probability, as a function of $|\eta|$, for electrons with $\pt>100\GeV$.
This separate measurement captures the effect of the charge consistency requirement being relaxed at high \pt (as described in Section~\ref{sec:Reconstruction}) on the charge misidentification rate.
We find values for this probability ranging from $10^{-4}$ for low \pt electrons in the central part of the detector to a few percent for high \pt electrons in the forward region of the detector.

The number of expected same-sign events due to charge misidentification is estimated by considering the total number of events passing the full selection but having oppositely charged leptons. 
These events are weighted by the charge misidentification probability parametrized as a function of $|\eta|$. 
The resulting  expected contribution of same-sign events due to charge misidentification is given in Table~\ref{tab:SummaryYields}.
A systematic uncertainty of 30\% for this background is assigned based on the variation of the charge misidentification probability observed between simulated Drell--Yan (DY) and $\ttbar$ MC events and also taking into account any potential \pt dependence for the statistically limited high-\pt region.

\subsubsection{Same-sign non-prompt background}
\label{sec:fakebkg}
In this category we consider non-prompt leptons that come from heavy-flavor decays, jets misidentified as leptons, decays in flight, or photon conversions. These contributions are estimated using the ``Tight-Loose'' method described in Ref.~\cite{susy2011} and used in our earlier publication~\cite{8TeVPaper}. This method relies on two definitions of leptons: ``tight'' and ``loose'', which are described in Section~\ref{sec:Reconstruction}. 

Any lepton passing either the tight or the loose selection can originate either from a prompt decay or from a non-prompt source, such as a heavy-flavor hadron, a misidentified hadron, or a photon converting to electrons. We refer to the former as ``prompt'' leptons and to the latter as ``fake'' leptons. The background is estimated
by using events with one or more loose leptons weighted by the ratios of the numbers of tight
leptons to the numbers of loose leptons expected for prompt and non-prompt leptons. The
ratio for prompt leptons is determined from observed DY events where the invariant mass of
the leptons is within 10\GeV of the $\PZ$ boson mass. We find a prompt rate of $0.873 \pm 0.001$ for electrons ($p_{\text{e}}$) and of $0.963 \pm 0.001$ for muons ($p_{\mu}$), where each reported error is the measurement's statistical error. The ``fake rate'', $f_{\ell}$, is defined as the probability that a fake lepton that passes the loose requirements will also pass the tight requirements. It is determined using a data sample enriched in non-prompt leptons. To reduce the contribution of leptons from \PW\ and $\PZ$ boson decays, exactly one loose lepton is required. We also require at least one jet with $ \pt > 30\GeV $ and $\Delta R$ \gt~1.0 relative to the lepton, $\MET < 25\GeV$, and $M_T < 25\GeV$, where $\Delta R$ is defined as $\sqrt{\smash[b]{(\Delta\phi)^{2}+(\Delta\eta)^{2}}}$, $\phi$ is the azimuthal angle measured in radians, and $M_T$ is the transverse mass of the lepton and \ptvecmiss. We also reject events if the invariant mass of the lepton and any jet is between 81 and 101\GeV. Fake rates of $0.286 \pm 0.003$ and $0.426 \pm 0.002$ are obtained for electrons and muons, respectively, where the reported errors are the statistical error on the measurement. The electron prompt and fake rates differ from those of muons because the electron identification and isolation criteria are more stringent than those for muons. The contribution of non-prompt leptons to the total background estimation is presented in Table~\ref{tab:SummaryYields}.

The systematic uncertainty in the estimation of backgrounds involving fake leptons is caused by the variations due to the flavor composition of the background (i.e. any dependence of the fake rates on the flavor source of the fake lepton), the level of closure in the method (studied in $\ttbar$ MC events), any potential dependence on kinematic parameters that alter the background composition (such as $\HTl$), as well as any potential dependence of the fake rate on $\eta$ or \pt.
The uncertaintiy due to these effects is found to be within 50\% and hence we assign a 50\% systematic uncertainty to the estimation of backgrounds due to fake leptons. 

\subsection{Event yields}
\begin{figure}[!htbp]
\centering
\includegraphics[width=0.45\textwidth]{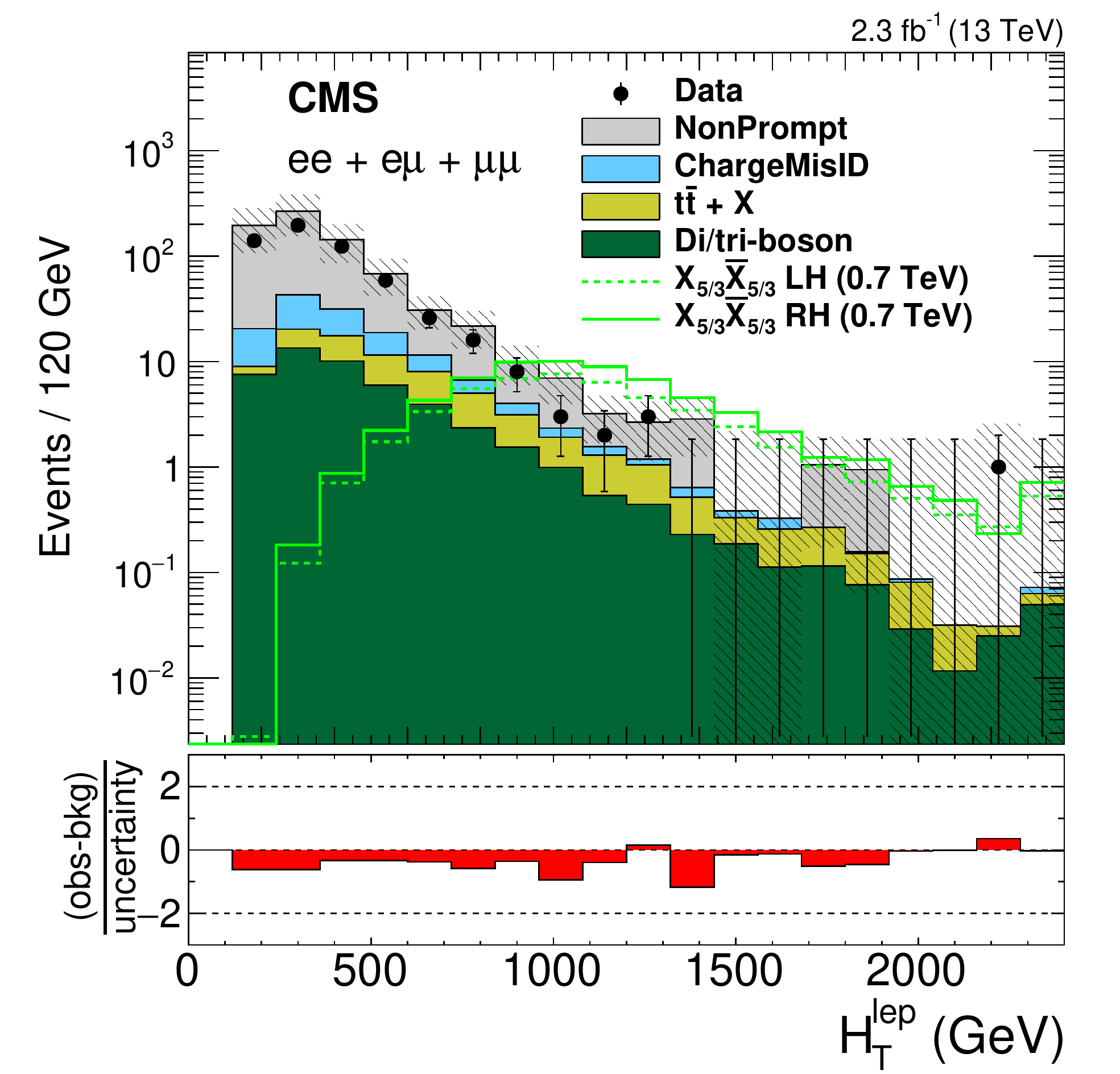}
\includegraphics[width=0.45\textwidth]{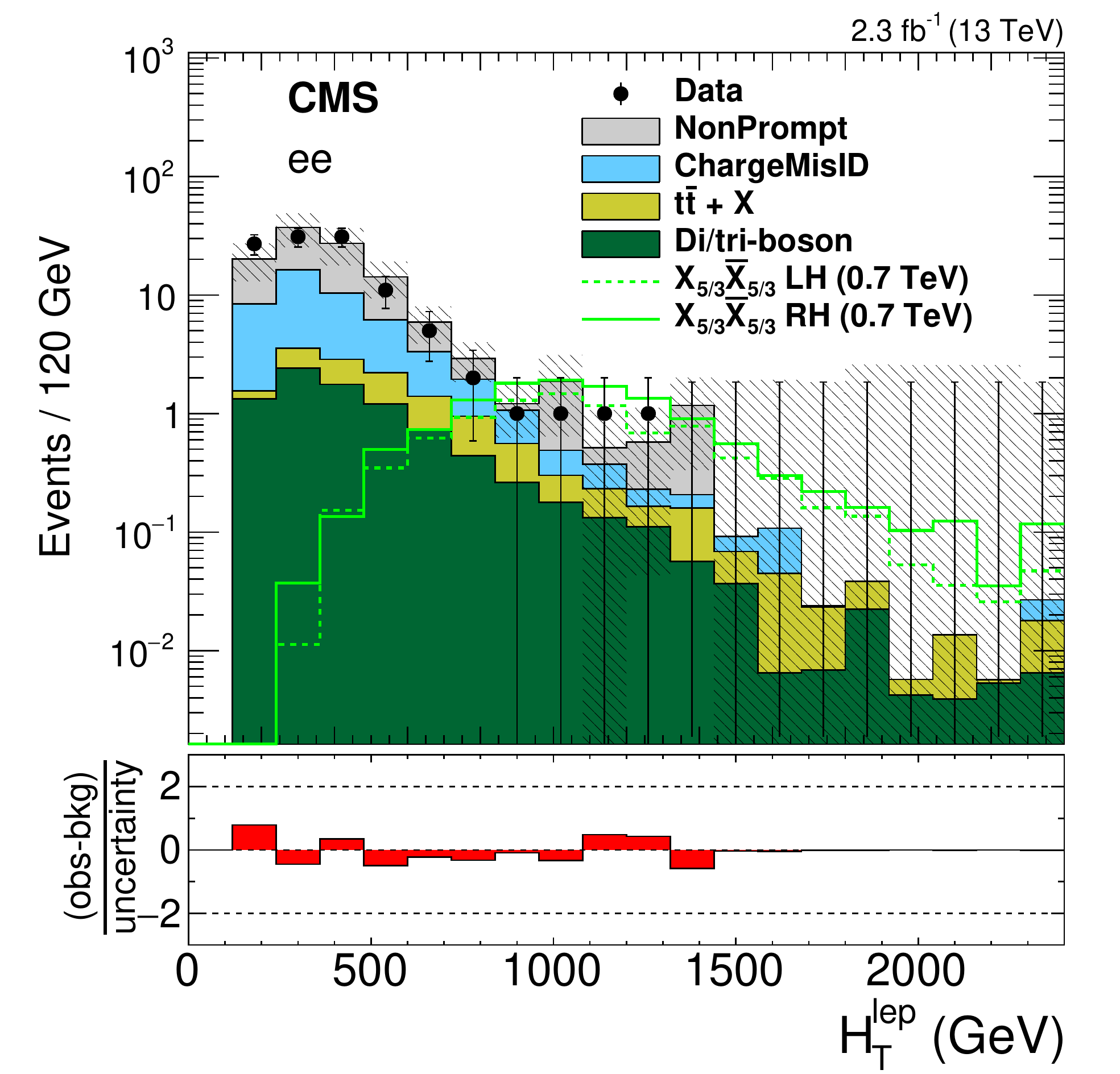}
\includegraphics[width=0.45\textwidth]{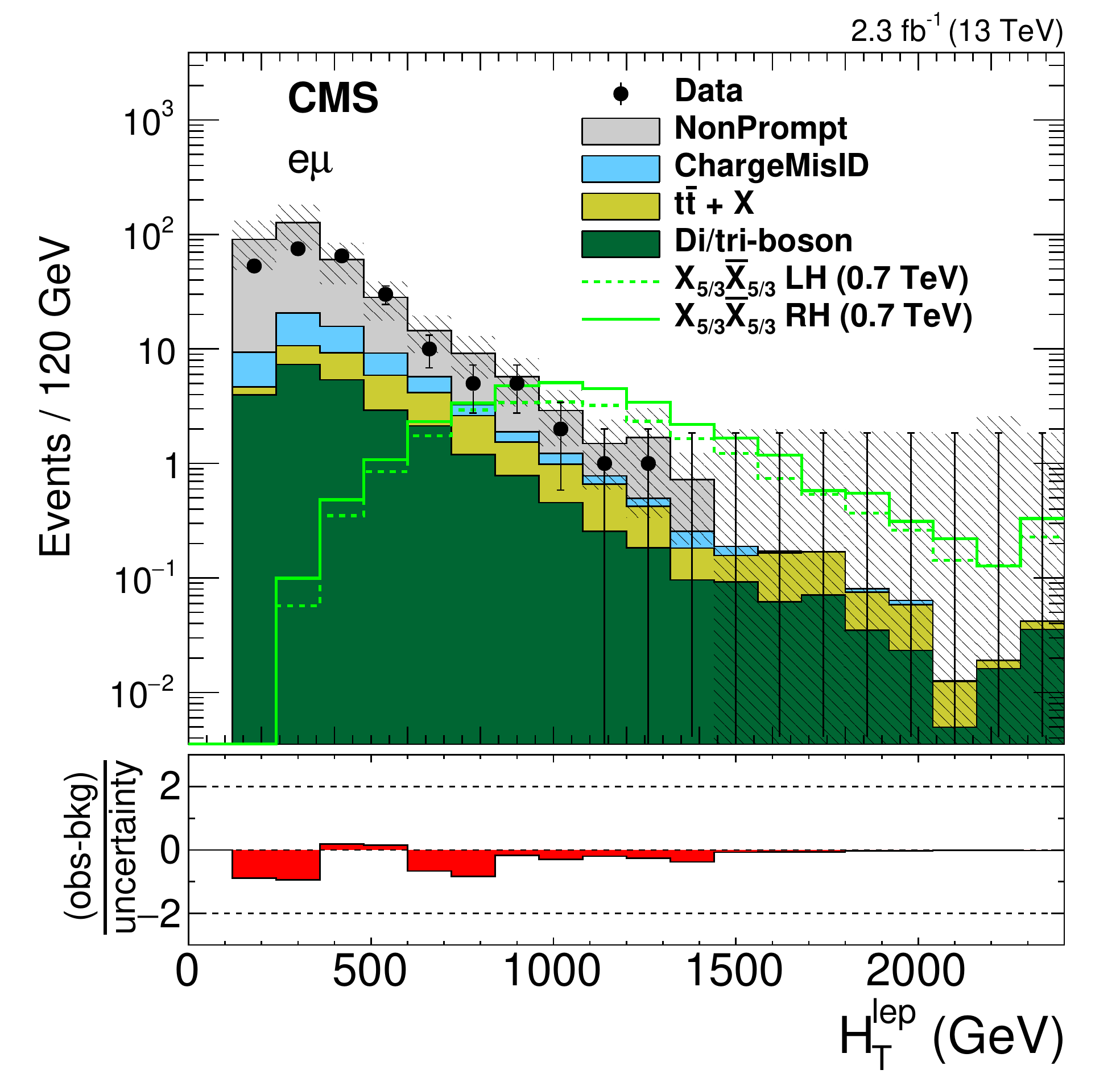}
\includegraphics[width=0.45\textwidth]{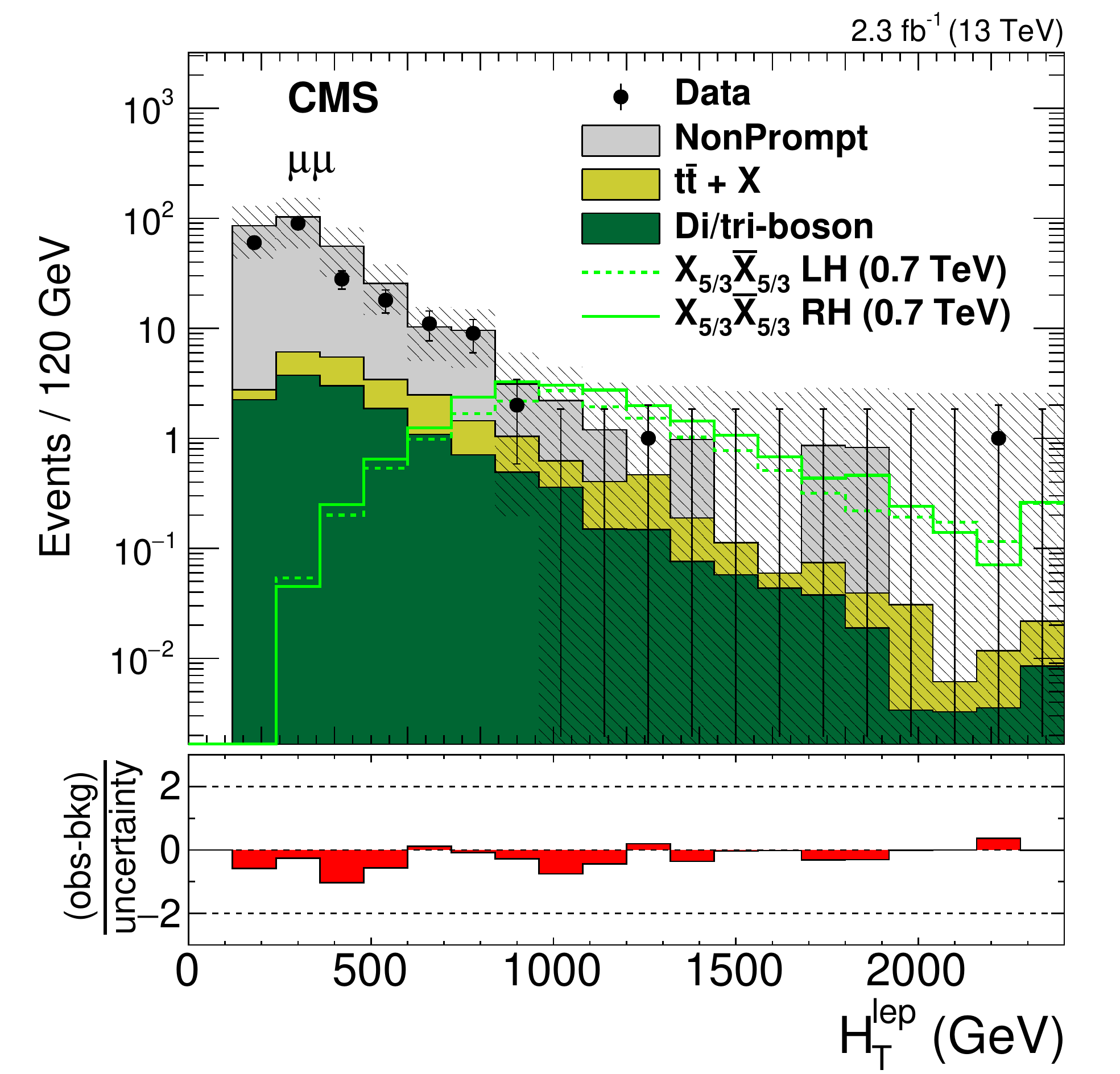}
\caption{The $\HTl$ distributions after the same-sign dilepton selection, \PZ/quarkonia lepton invariant mass vetoes, and the requirement of at least two AK4 jets in the event. The hatched area shows the combined systematic and statistical uncertainty in the background prediction for each bin. The lower panel in all plots shows the difference between the observed and the predicted numbers of events divided by the total uncertainty. The total uncertainty is calculated as the sum in quadrature of the statistical uncertainty in the observed measurement and the uncertainty in the background, including both statistical and systematic components. Also shown are the distributions for a 700\GeV \xft with right-handed (solid line) and left-handed (dashed line) couplings to W bosons.}
\label{fig:AK4HT}
\end{figure}

Figure~\ref{fig:AK4HT} shows the $\HTl$ distributions after applying the quarkonia veto, associated $\PZ$ boson veto, primary $\PZ$ boson veto, and a requirement of at least two AK4 jets in the event. These distributions are for illustrative purposes only: the full selection is not applied because of the limited number of events. The uncertainty bands in the upper and lower panels of each plot include both statistical and systematic uncertainties.

The total number of expected background events are reported in Table~\ref{tab:SummaryYields}, together with the numbers of observed and expected events for a right-handed \xft with a mass of 800\GeV.
In total four events are observed, which is consistent with the predicted background, taking its uncertainty into account.
\begin{table} [!htbp]
\centering
\topcaption{Summary of background yields from SM processes with two same-sign prompt leptons (SSP MC), same-sign non-prompt leptons (NonPrompt), and opposite-sign prompt leptons (ChargeMisID), as well as observed data events  after the full analysis selection for the same-sign dilepton channel, with an integrated luminosity of 2.3\fbinv. Also shown are the numbers of expected events for a right-handed \xft with a mass of 800\GeV. The uncertainties include both statistical and systematic components, as discussed in Section~\ref{sec:Systematics}.}
\resizebox{\columnwidth}{!}{%
\begin{tabular}{c|ccc|c|c|c}\hline
Channel & SSP MC&NonPrompt&ChargeMisID& Total background & 800\GeV \xft&Observed\\
\hline
Dielectron&$0.7\pm0.1$&$1.2\pm1.0$&$0.2\pm0.1$ &$ 2.1\pm1.0$&$3.2\pm0.3$ & 1\\
Electron-muon&$1.7\pm0.2$&$2.6\pm2.0$&$0.3\pm0.1$ &$ 4.6\pm2.0$&$9.1\pm0.7$ & 1\\
Dimuon&$1.2\pm0.2$&$4.6\pm3.0$&$0.0\pm0.0$ &$ 5.8\pm3.0$&$5.6\pm0.4 $& 2\\
\hline
Total &$3.6\pm0.4$&$8.4\pm5.0$&$0.5\pm0.2$ &$ 12.5\pm5.0\x$&$17.9 \pm1.3\x$& 4\\
\hline
\end{tabular}
}
\label{tab:SummaryYields}
\end{table}

\section{Single-lepton final state}

The search for \xft in the single-lepton final state 
targets events where one of the $\PW$ bosons decays into a lepton and a neutrino, 
while the other three $\PW$ bosons decay hadronically. The SM background processes 
leading to a similar final state can be grouped into three categories: 
top quark, electroweak and QCD multijet backgrounds. 
The ``top quark background'' group, labeled ``TOP'', is dominated by $\ttbar$ pair production 
and also includes single top quark production processes and the rare SM processes $\ttbar\PW$ 
and $\ttbar\PZ$ (the $\ttbar\PH$ contribution is negligible). 
The ``electroweak background'' group, labeled ``EWK'', is dominated by 
$\PW$+jets production, and includes the DY and diboson ($\PW\PW$, $\PW\PZ$, $\PZ\PZ$) contributions. 

A preselection of events is made by requiring exactly one lepton with $ \pt > 50\GeV$ that also passes the tight identification and isolation requirements described in Section~\ref{sec:Reconstruction}. 
Events containing any additional loose lepton with $ \pt > 10\GeV$ are ignored.

Because of the significant amount of jet activity in the final state for a potential signal, 
we require at least three jets, 
where the \pt of the leading jet is greater than 200\GeV and that of the subleading jet is greater than 90\GeV.
To remove the residual multijet events in which jets overlap with the lepton, 
an additional selection criterion is imposed by requiring that 
the lepton and the closest jet either be separated by $\Delta R$($\ell$, closest jet)~\gt~0.4, 
or the magnitude of the lepton \pt perpendicular to the jet axis be larger than 40\GeV.
In order to suppress the multijet background contribution, a large missing transverse energy
requirement, $ \MET > 100\GeV$, is imposed.

A discriminant produced by the combined secondary vertex (CSVv2) algorithm~\cite{CMS-PAS-BTV-15-001}
is used to identify jets that are likely to have originated from the production of a bottom quark. 
At the discriminant value used to select b-tagged jets, 
the algorithm has a single-jet signal efficiency of ${\sim}65\%$ and a light quark mistag efficiency of only ${\sim}1\%$. 
We require at least one of the jets in each event to be $\PQb$ tagged.

\begin{figure}[hbtp]
\begin{center}
\includegraphics[width=0.49\textwidth]{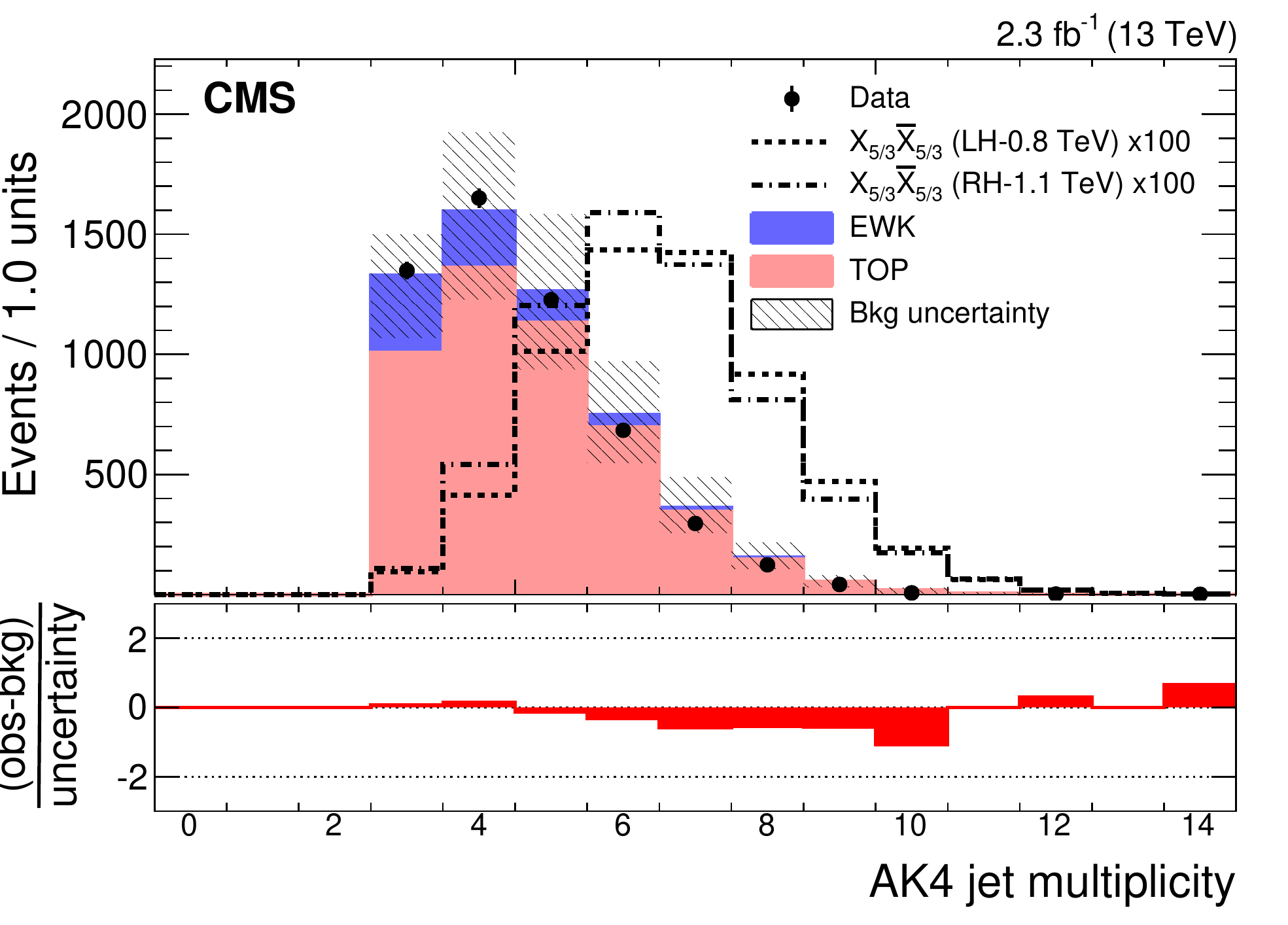}
\includegraphics[width=0.49\textwidth]{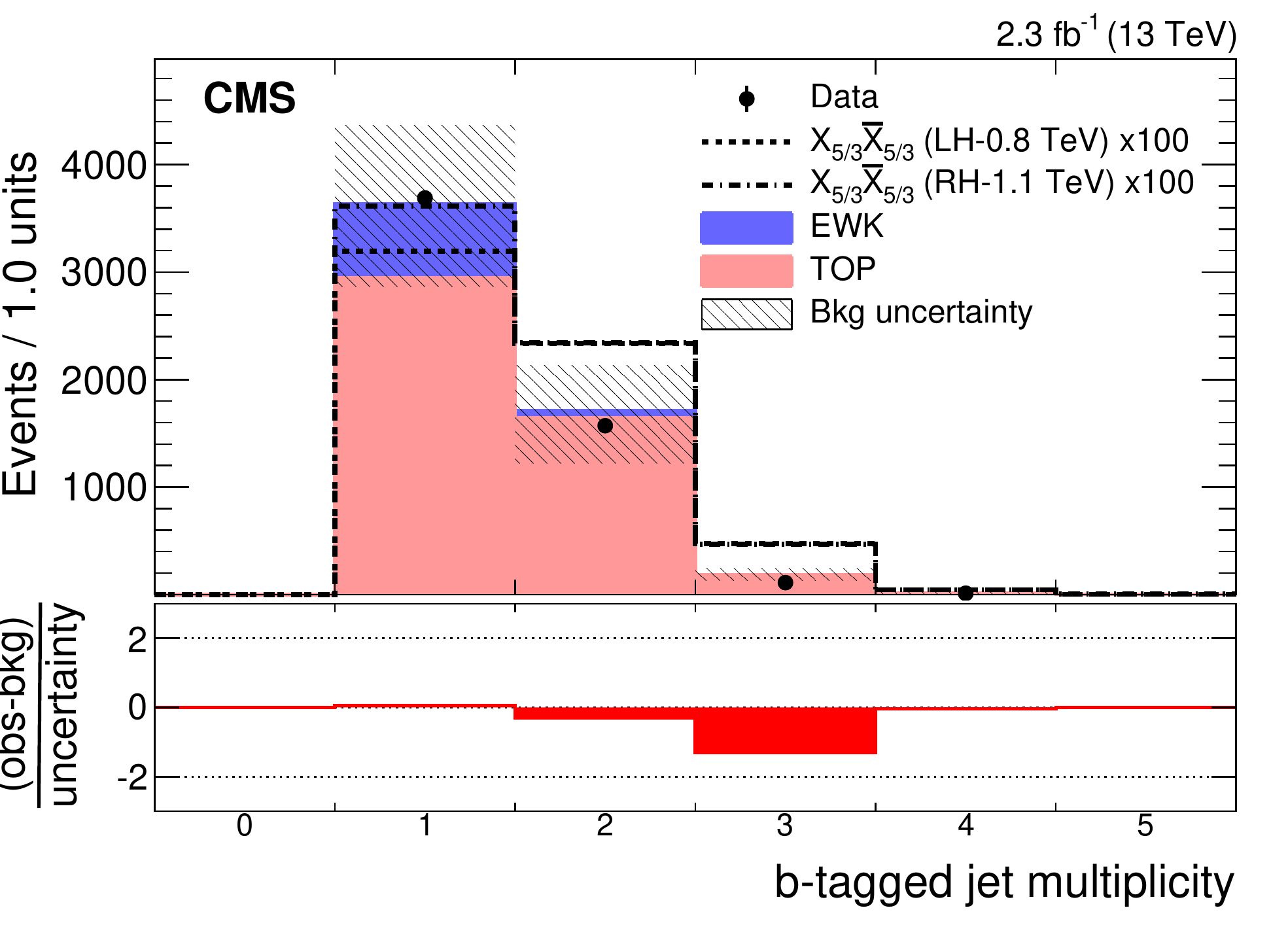}
\includegraphics[width=0.49\textwidth]{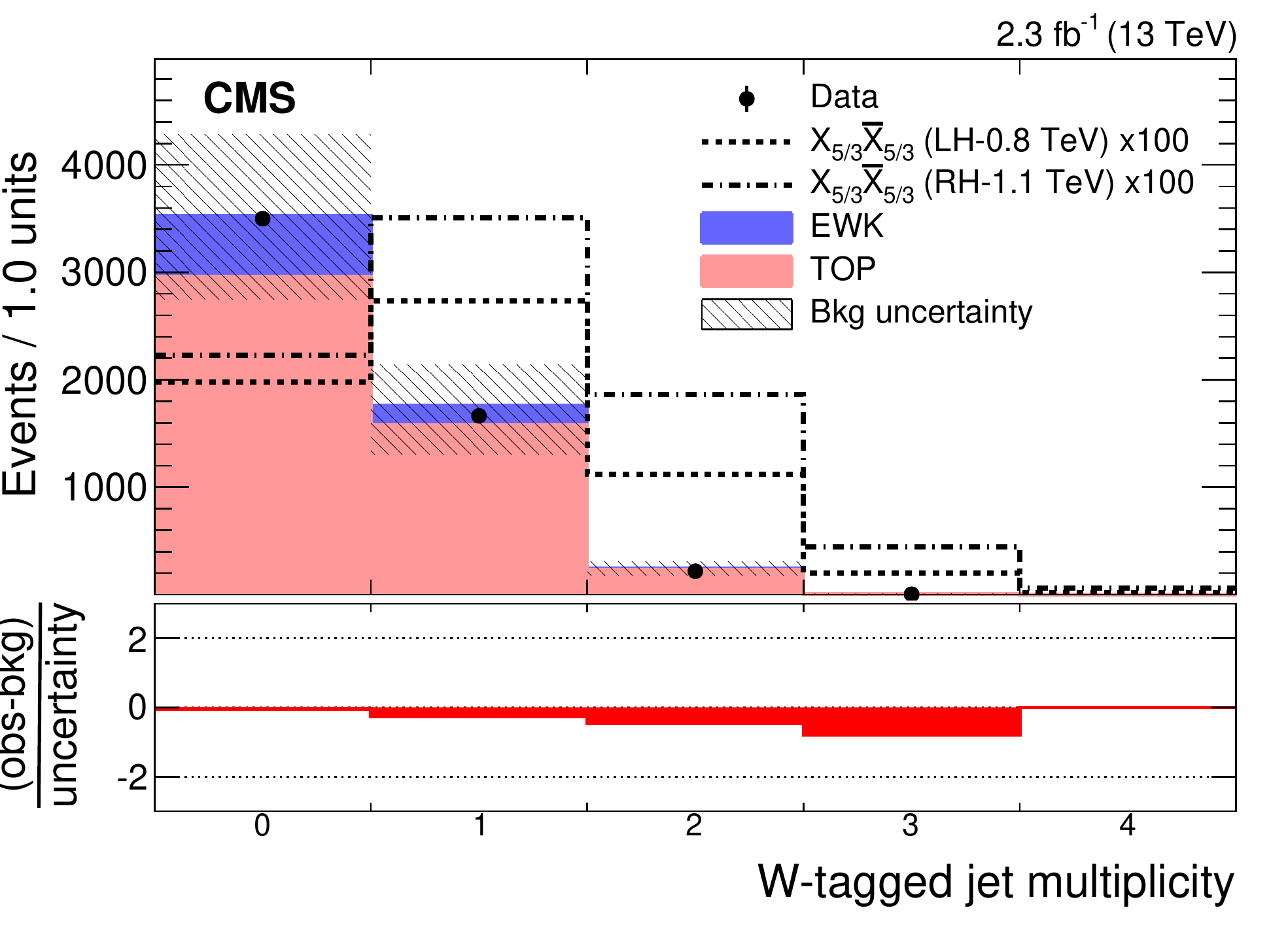}
\includegraphics[width=0.49\textwidth]{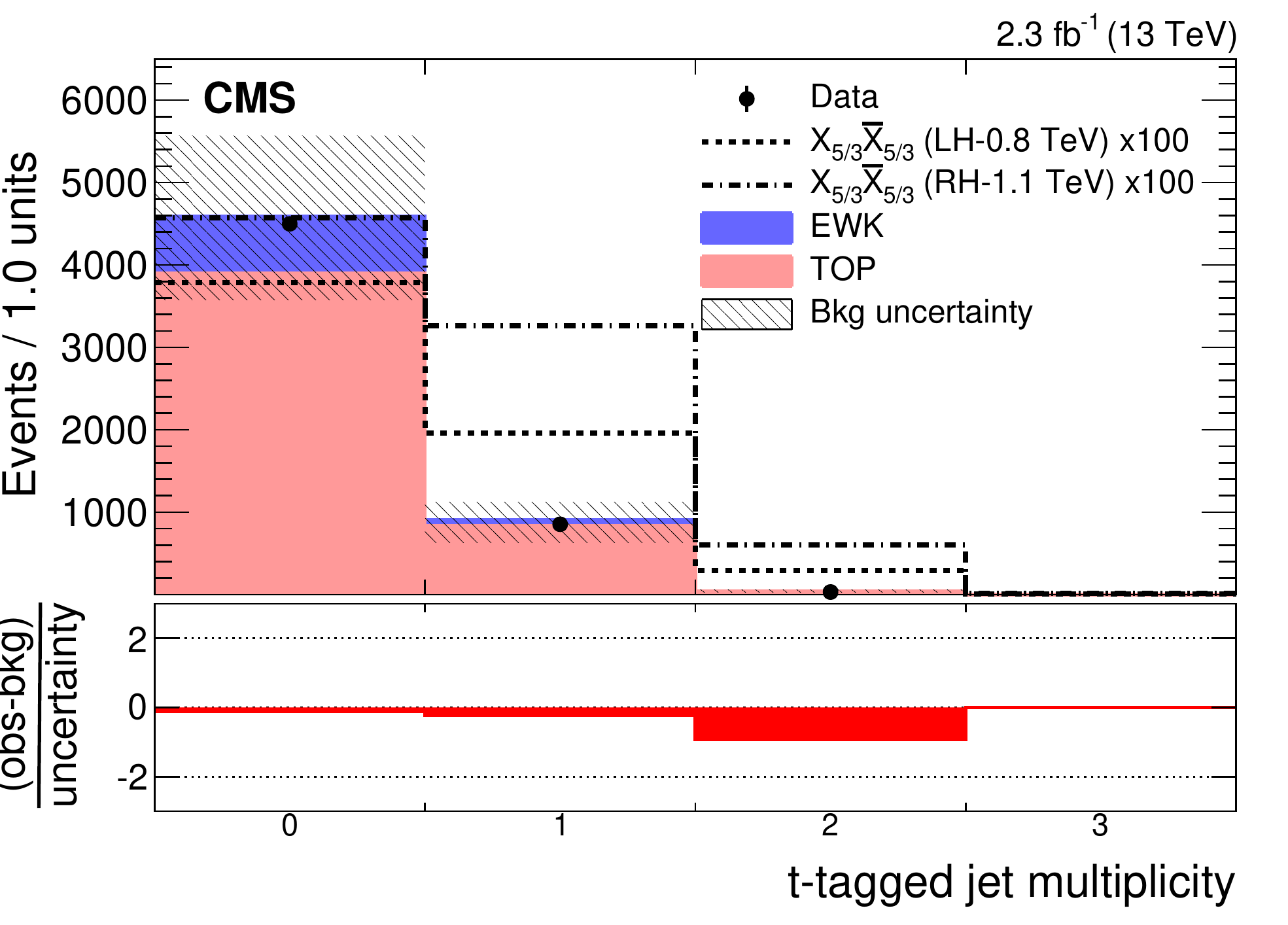}
\caption{Distributions of the number of AK4 jets (upper left), the numbers of $\PQb$-tagged (upper right), $\PW$-tagged (lower left), and $\PQt$-tagged jets (lower right) in data and simulation for combined electron and muon event samples, at the preselection level. 
The lower panel in all plots shows the difference between the observed and the predicted numbers of events divided by the total uncertainty. 
The total uncertainty is calculated as the sum in quadrature of the statistical uncertainty in the observed measurement and the uncertainty in the background, including both statistical and systematic components.
Also shown are the distributions of representative signal events, which are scaled by a factor of 100.}
\label{fig:preselTags}
\end{center}
\end{figure}

Decay products of heavy particles such as \xft can have large Lorentz boosts, 
and their subsequent decay products can merge into a single jet.
The substructure of these jets is explored using larger-radius jets, reconstructed 
with an anti-$k_{t}$ distance parameter of $0.8$ (AK8),
in order to identify merged jets that are likely to originate from a $\PW$ boson
or a top quark~\cite{tTAGSFS}.
The ``$N$-subjettiness''~\cite{NSUBJETS} algorithm measures the likelihood of a jet 
having $N$ subjets ($N=\;$1, 2, 3, etc). Jet grooming techniques are used to remove soft 
jet constituents so that the mass of the hard constituents can be measured more precisely. 
The ``pruning''~\cite{PRUNING} and ``soft-drop''~\cite{SOFTDROP} algorithms are 
used to identify boosted hadronic $\PW$ boson decays and 
boosted hadronic top quark decays, respectively. 
The $\PW$-tagged jets are required to have $ \pt > 200\GeV$, $|\eta|<2.4$, 
pruned mass between 65 and 105\GeV, and the ratio of $N$-subjettiness variables~\cite{NSUBJETS} $\tau_2/\tau_1 < 0.6$,
which ensures that the $\PW$-tagged jets are more likely to have two subjets than one subjet.
The pruned jet mass scale and resolution, along with the efficiency of the $\tau_2/\tau_1$ selection,
are compared between data and simulation in a control region dominated by $\ttbar$ events with boosted hadronic $\PW$ boson decays 
and scale factors are applied in the simulation
to match them with the performance found in data.
The $\PQt$-tagged jets are required to have $ \pt > 400\GeV$, $|\eta|<2.4$, 
soft-drop mass between 110 and 210\GeV, and the ratio of $N$-subjettiness variables $\tau_3/\tau_2 < 0.69$,
which ensures that the $\PQt$-tagged jets are more likely to have three subjets than two subjets.  
Figure~\ref{fig:preselTags} shows the number of AK4 jets, as well as the numbers of $\PQt$-, $\PW$-, 
and $\PQb$-tagged jets. The figure also shows that, at this level of the selection, 
the sample is largely dominated by top quark events, with some contribution from electroweak processes; 
the contribution from QCD multijet processes is negligible.

In a second step, the selections on the lepton \pt, \MET, jet \pt, number of AK4 jets, 
and on the distance between the lepton and the subleading jet, $\Delta R$($\ell$, $j_{2}$),
are optimized in a procedure that minimizes the upper limit on the \xft cross section 
expected in the absence of a signal. This procedure was also cross checked with an alternative method 
that maximizes the expected significance and similar selection requirements have been found.
The final selection demands, in addition to the preselection requirements listed earlier, 
the presence of at least four jets, the lepton $ \pt > 80\GeV$, and $\Delta R$($\ell$, $j_{2}$)~\gt~1.

\begin{figure}[hbtp]
\begin{center}
\includegraphics[width=0.49\textwidth]{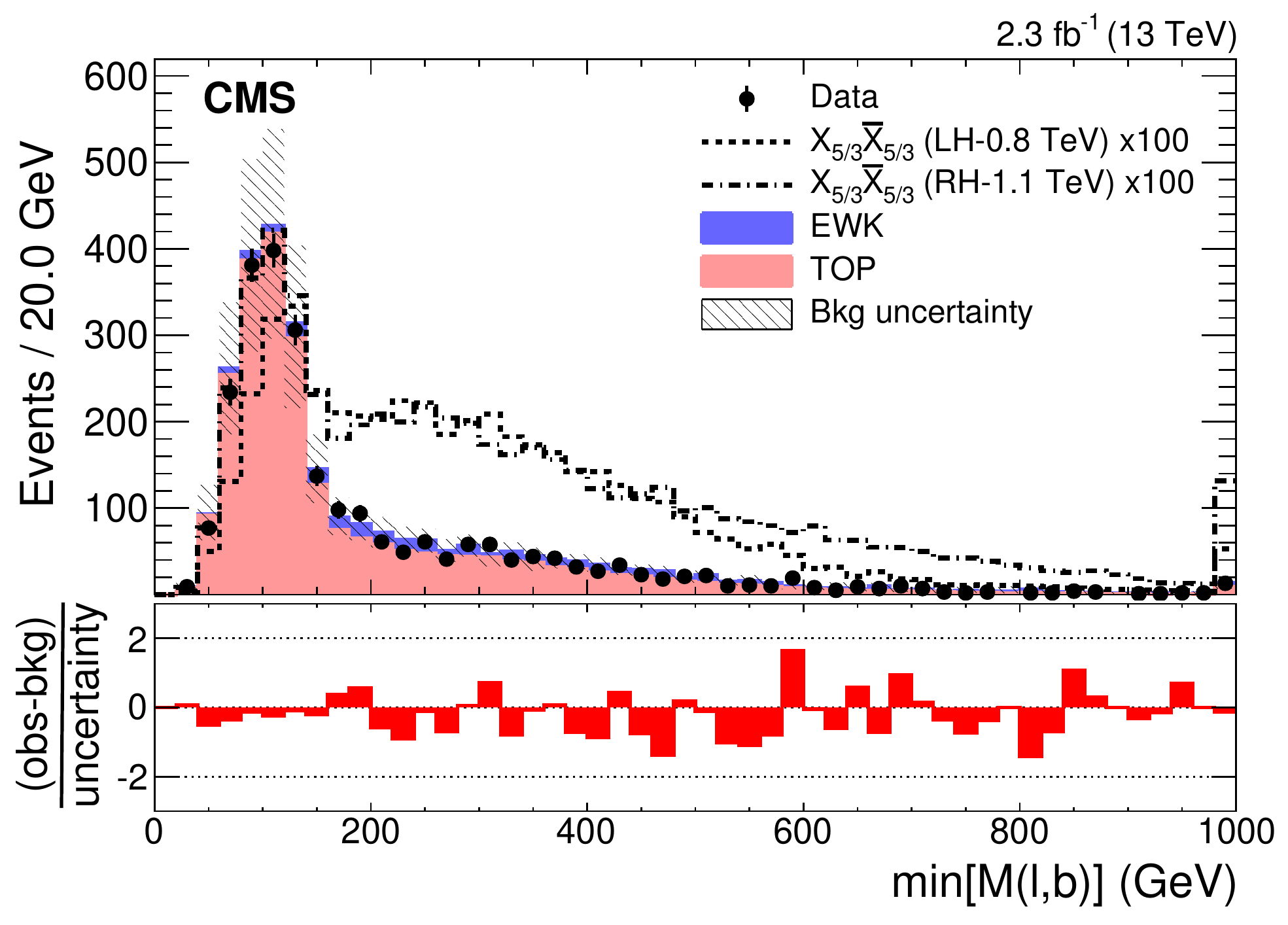}
\includegraphics[width=0.49\textwidth]{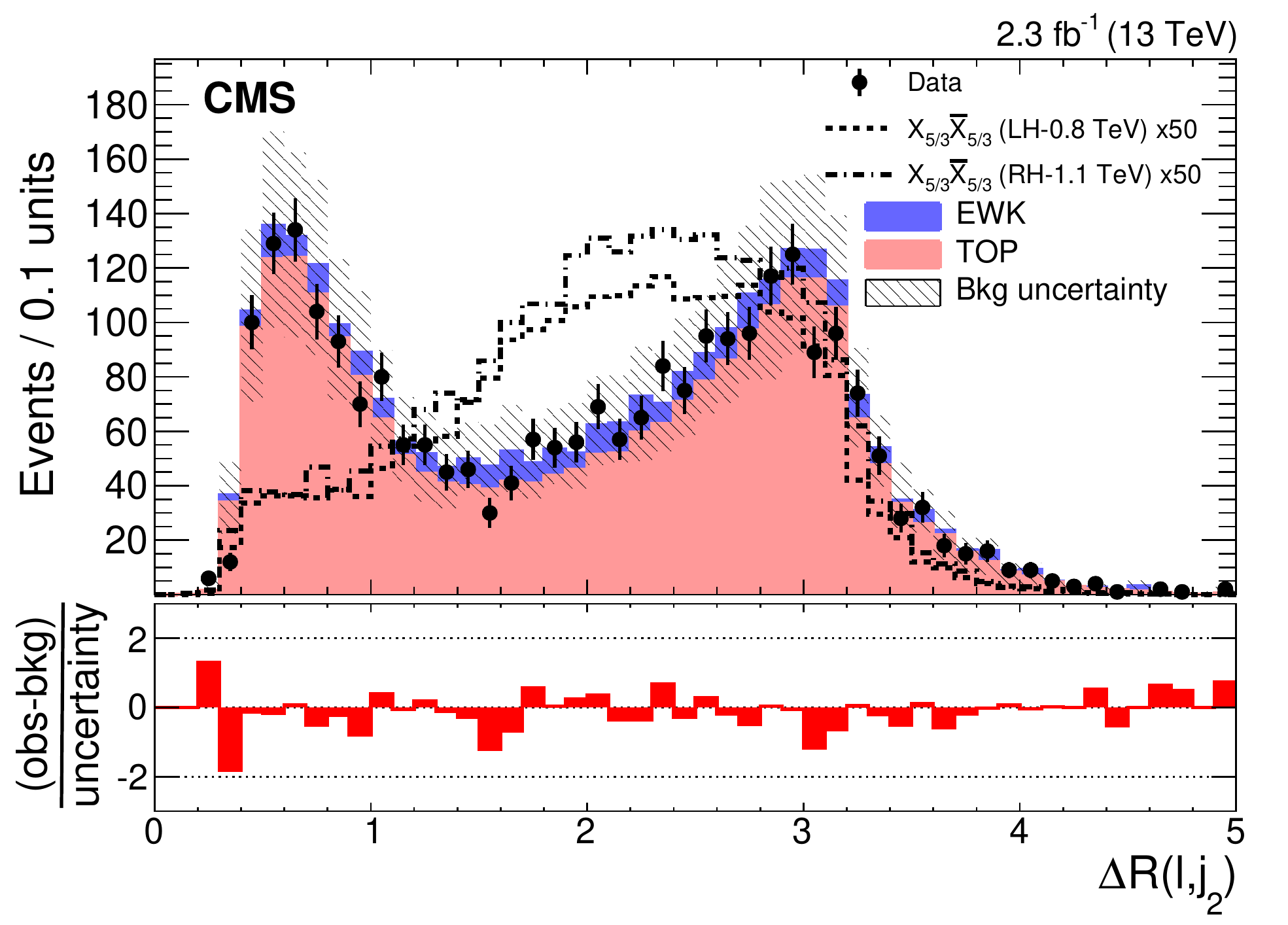}
\caption{Distributions of min[M($\ell,\,\PQb$)] (left) and  $\Delta R$($\ell$, $j_{2}$) (right) in data and simulation for selected events with at least four jets and lepton $ \pt > 80\GeV$. 
The lower panel in all plots shows the difference between the observed and the predicted numbers of events divided by the total uncertainty. 
The total uncertainty is calculated as the sum in quadrature of the statistical uncertainty in the observed measurement and the uncertainty in the background, including both statistical and systematic components.
Also shown are the min[M($\ell$,\,$\PQb$)] ($\Delta R$($\ell$, $j_{2}$)) distributions of representative signal events, 
which are scaled by a factor of 100 (50) so that the shape differences between signal and background are visible.}
\label{fig:sel4jets}
\end{center}
\end{figure}

The mass constructed from the lepton and $\PQb$-tagged jet, labeled M($\ell$,\,$\PQb$), 
provides good discrimination between signal and background. 
In case more than one $\PQb$-tagged jet is found in the event,
the one that leads to the smallest M($\ell$,\,$\PQb$) defines the discriminating variable, min[M($\ell$,\,$\PQb$)],
which is used in the analysis to extract or constrain the signal. The distribution of 
min[M($\ell$,\,$\PQb$)] is shown in Fig.~\ref{fig:sel4jets}, together with the distance between the lepton 
and the subleading jet in the event, $\Delta R$($\ell$, $j_{2}$),
for events passing the final selection criteria, except for the requirement on $\Delta R$($\ell$, $j_{2}$).
The distribution of min[M($\ell$,\,$\PQb$)] for the background, dominated by $\ttbar$ events, 
features a sharp drop around 150\GeV, since, for such events,
this variable represents the visible mass of the top quark in the detector.
The $\Delta R$($\ell$, $j_{2}$) variable shows that the subleading jets populate 
both the same and opposite hemisphere relative to the lepton in the background events, 
whereas in the \xft signal events, the subleading jet is usually opposite to the lepton.
This is used in the final selection to further suppress the background contribution in the signal 
region as well as to reduce the signal contamination in the control region, as discussed in the following section.

\subsection{Background modeling}
\label{sec:background}
In the single-lepton final state analysis, all the SM background processes are estimated
using simulation. To cross check the background modeling, we consider two control regions
to study the two dominant background processes in this analysis:
one enriched in $\ttbar$ events, and the other enriched in $\PW$+jets events. 
In order to define these control regions, events are selected by imposing the same
requirements as for the final selection apart from the $\Delta R$($\ell$, $j_{2}$) and the $\PQb$ tagging requirements.
The selection on $\Delta R$($\ell$, $j_{2}$) is inverted, requiring this variable to be less than~1.

The $\ttbar$ background control region is then defined by selecting events that have 
${\ge}$1 $\PQb$-tagged jets, while the $\PW$+jets control region is
obtained by requiring the presence of 0 $\PQb$-tagged jets. For the $\PW$+jets sample, 
owing to the 0 $\PQb$-tagged jet requirement, 
we use each and every selected jet in the event as a $\PQb$-jet candidate to obtain 
the mass discriminant, and denote it as min[M($\ell$,\,jet)].

\begin{figure}[hbtp]
\begin{center}
\includegraphics[width=0.49\textwidth]{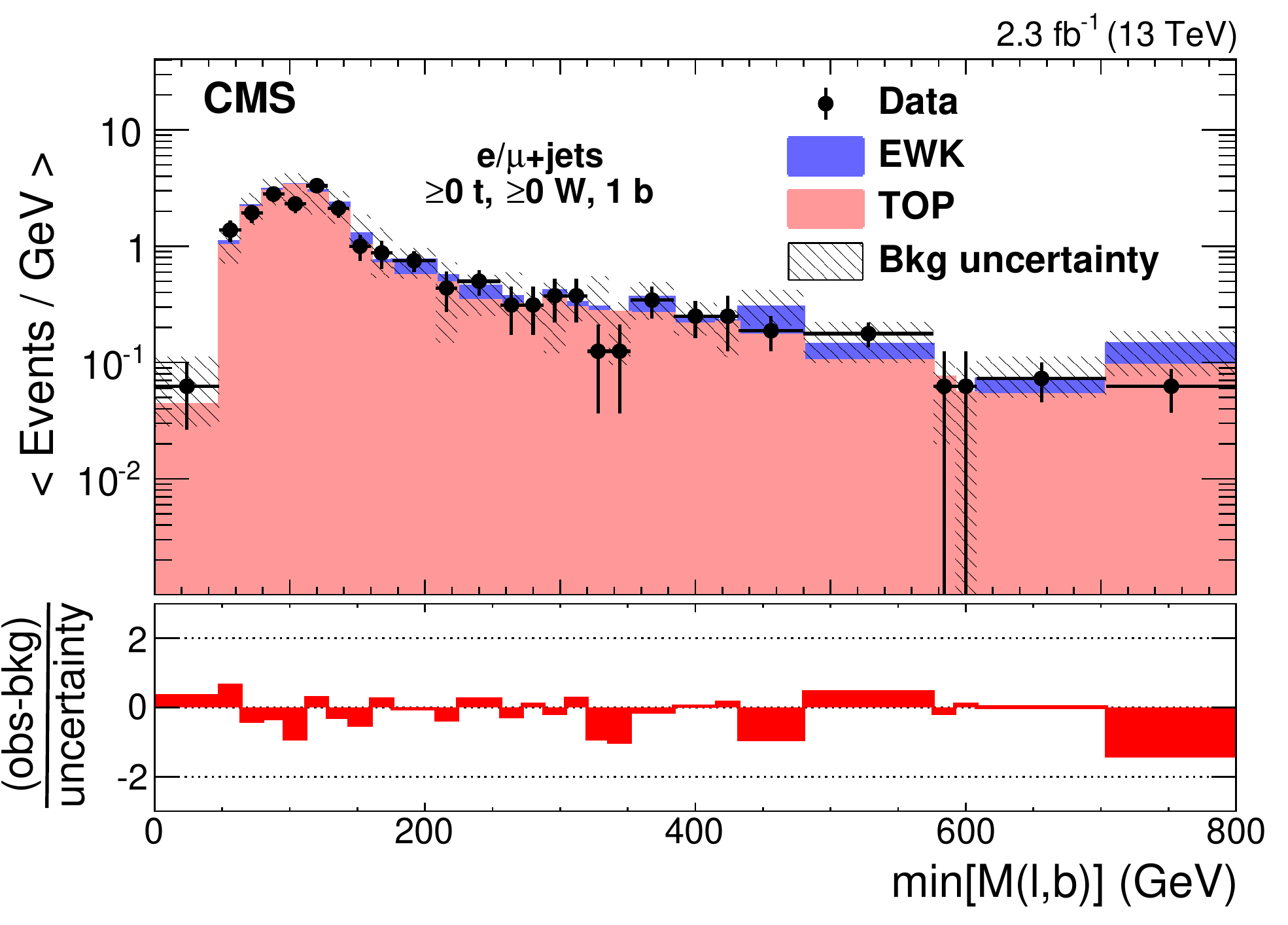}
\includegraphics[width=0.49\textwidth]{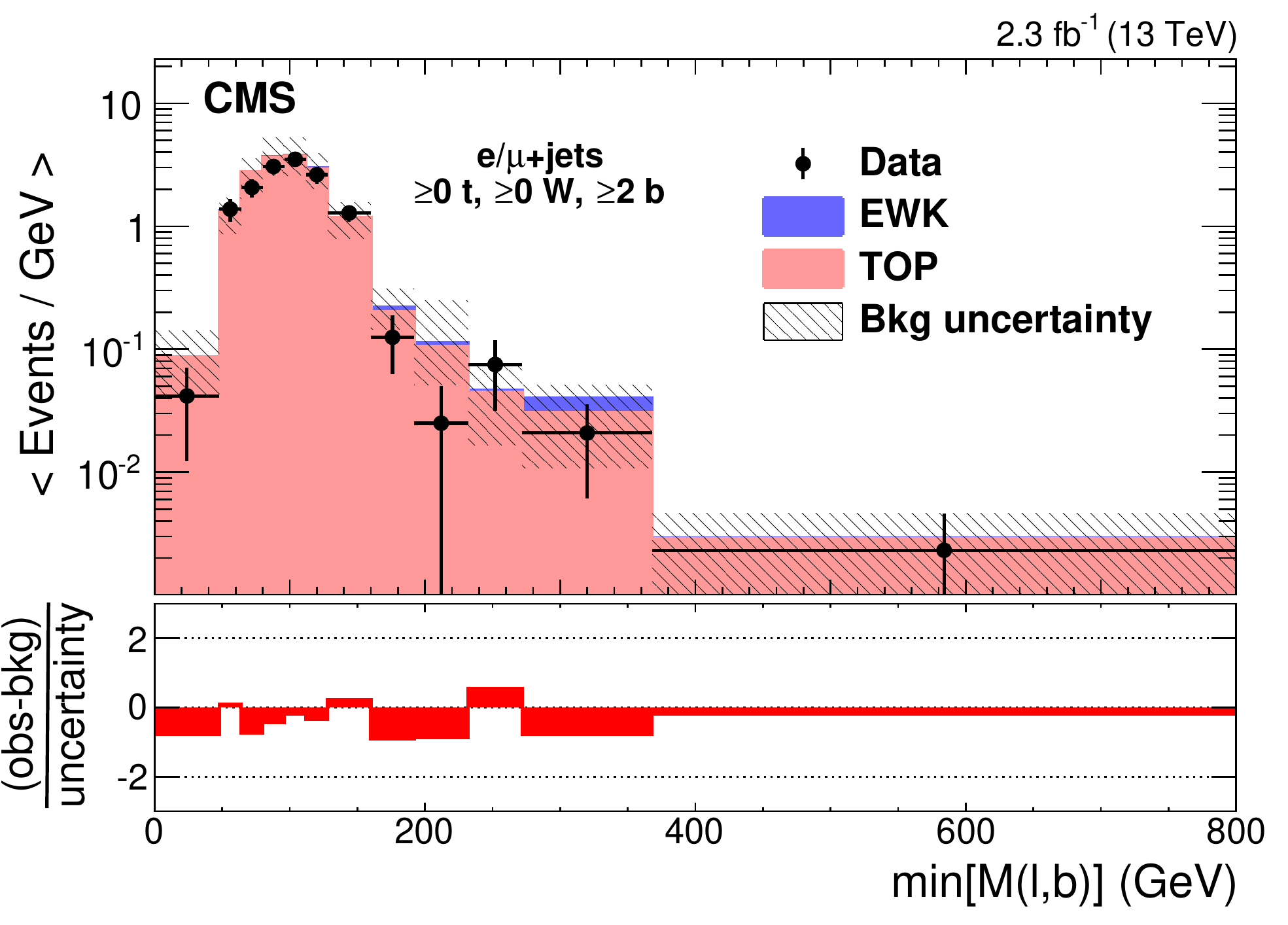}
\includegraphics[width=0.49\textwidth]{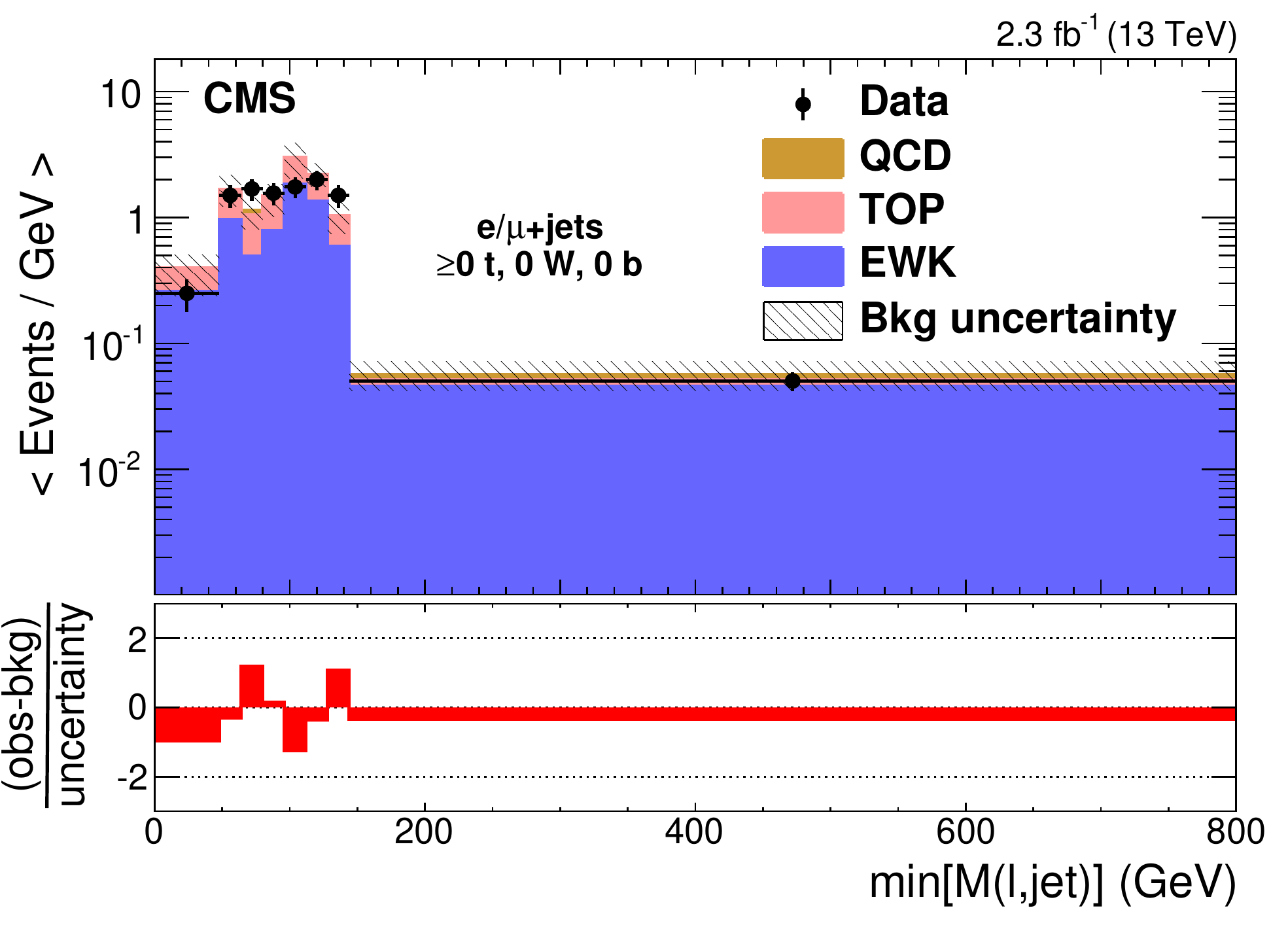}
\includegraphics[width=0.49\textwidth]{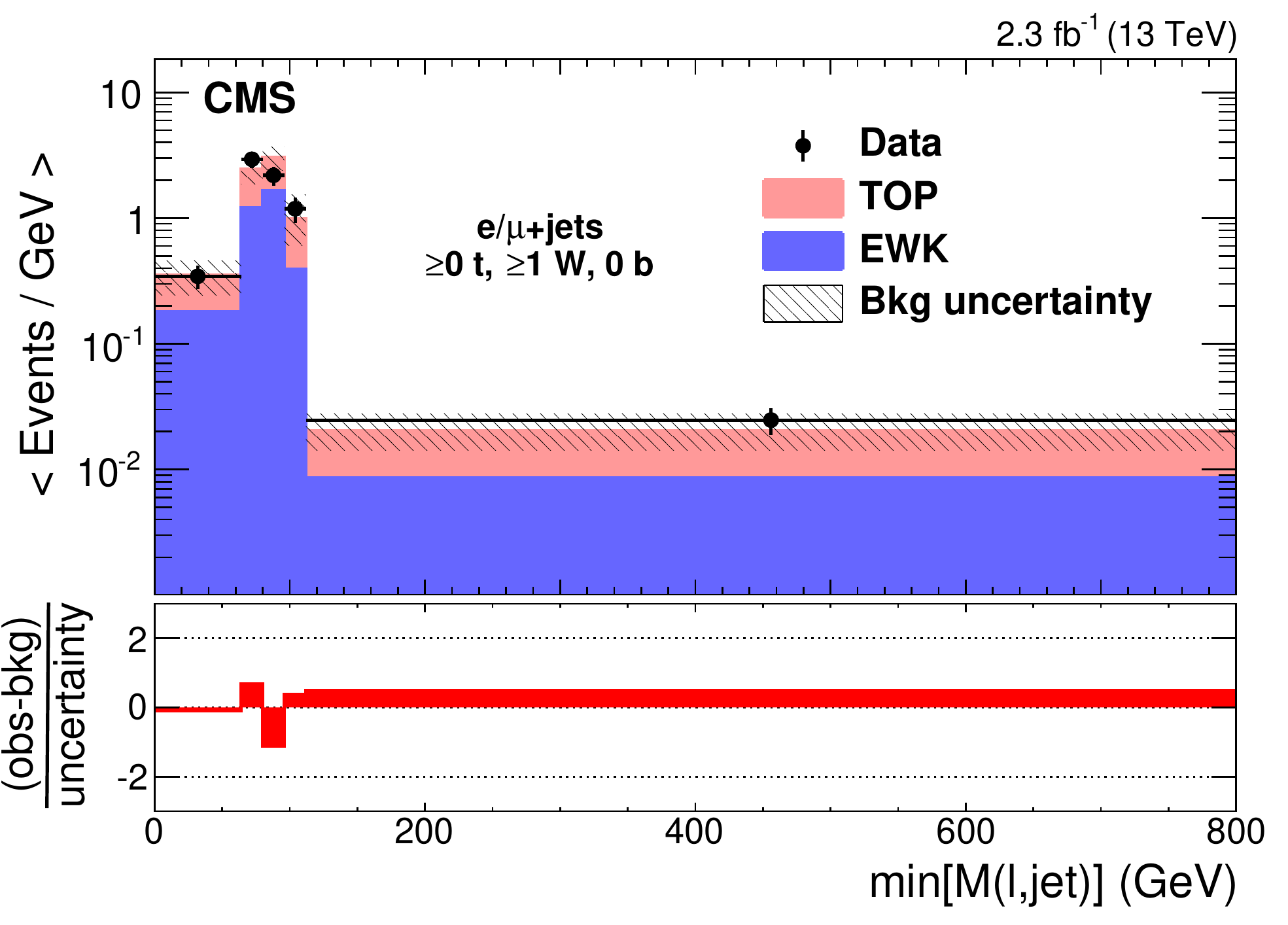}
\caption{Distributions of min[M($\ell$,\,$\PQb$)] in the $\ttbar$ control region, for 1 $\PQb$-tagged jet (upper left) and 
$\ge$2 $\PQb$-tagged jets (upper right) categories, and of min[M($\ell$,\,jet)] in the $\PW$+jets control region, 
for 0 $\PW$-tagged (lower left) and  $\ge$1 $\PW$-tagged jet (lower right) categories for 
combined electron and muon event samples.
The horizontal bars on the data points indicate the bin widths. 
The lower panel in all plots shows the difference between the observed and the predicted numbers of events divided by the total uncertainty. 
The total uncertainty is calculated as the sum in quadrature of the statistical uncertainty in the observed measurement and the uncertainty in the background, including both statistical and systematic components.
A small QCD multijet contribution is displayed in the bottom left plot; in all other distributions, 
it is less than 0.5\% and is not shown.}
\label{fig:bkgCR}
\end{center}
\end{figure}

In the $\ttbar$ control region, the events are split into two categories, one with exactly 1 $\PQb$-tagged jet, 
and the other with two or more $\PQb$-tagged jets. 
For the $\PW$+jets control region, we also define two categories of events, 
but now based on the number of $\PW$-tagged jets: 0 $\PW$-tagged, or 1 or more $\PW$-tagged jets.  
Figure~\ref{fig:bkgCR} shows the min[M($\ell$,\,$\PQb$)] (min[M($\ell$,\,jet)]) distributions 
in the $\ttbar$ ($\PW$+jets) control region.
The comparison of the observed and the predicted yields in the control regions for
each tagging category is used as a closure test for background modeling.
In both control regions, the background predictions based on simulation show good agreement with data,
and any deviation from unity of the ratio between data and simulation is well within the combined uncertainties. 

\subsection{Event yields}
\begin{figure}[hbtp]
\begin{center}
\includegraphics[width=0.49\textwidth]{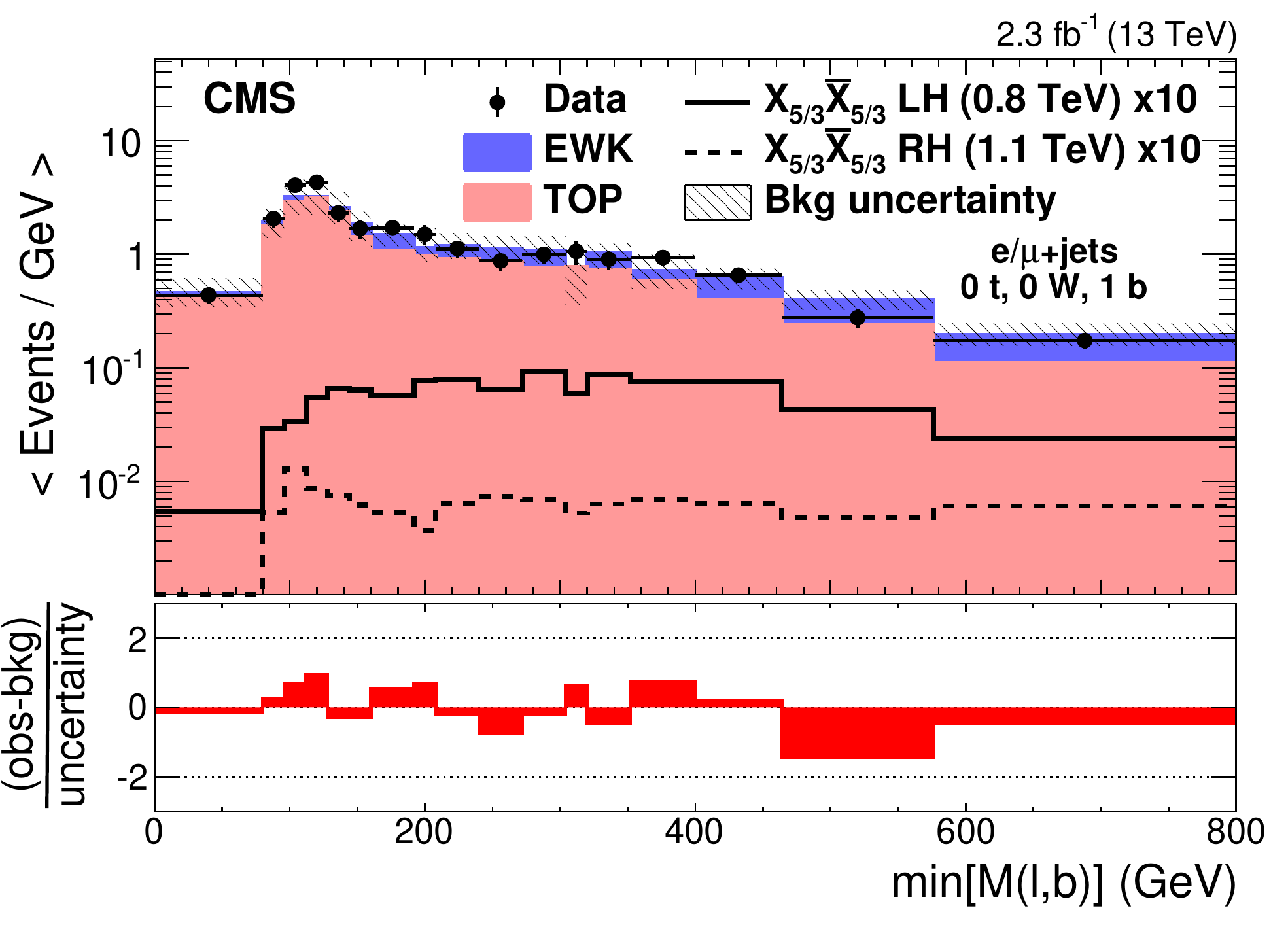}
\includegraphics[width=0.49\textwidth]{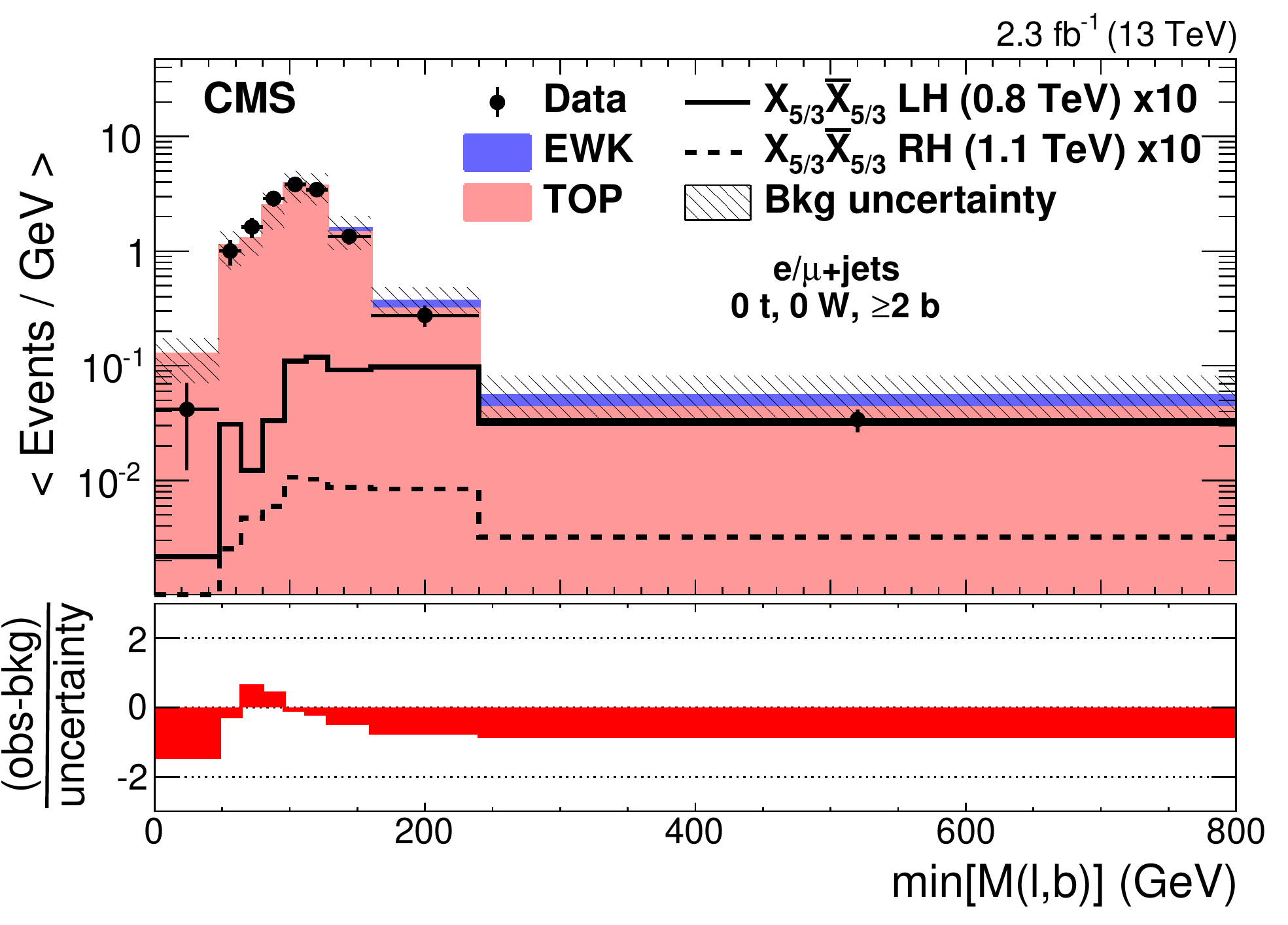}
\includegraphics[width=0.49\textwidth]{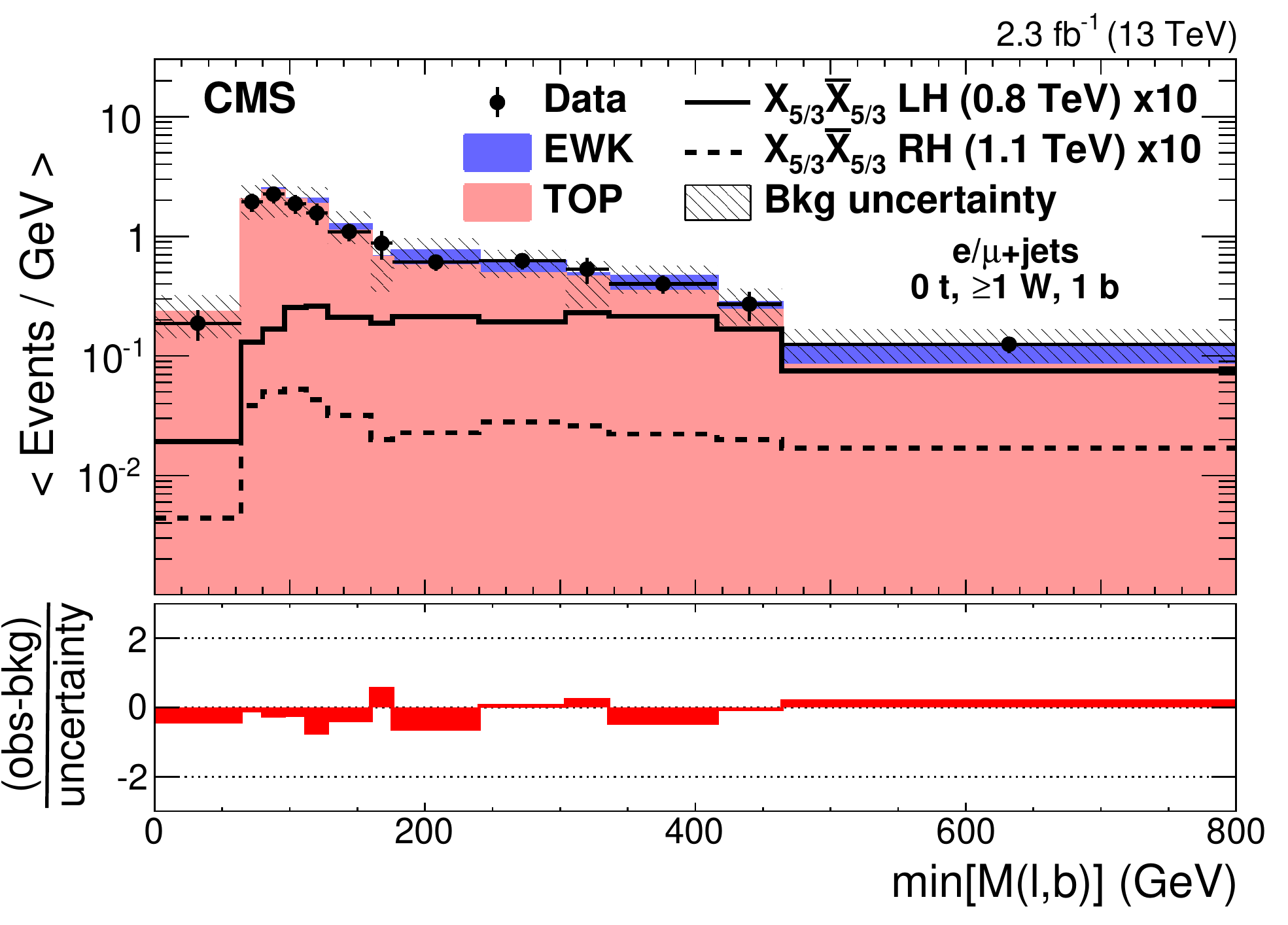}
\includegraphics[width=0.49\textwidth]{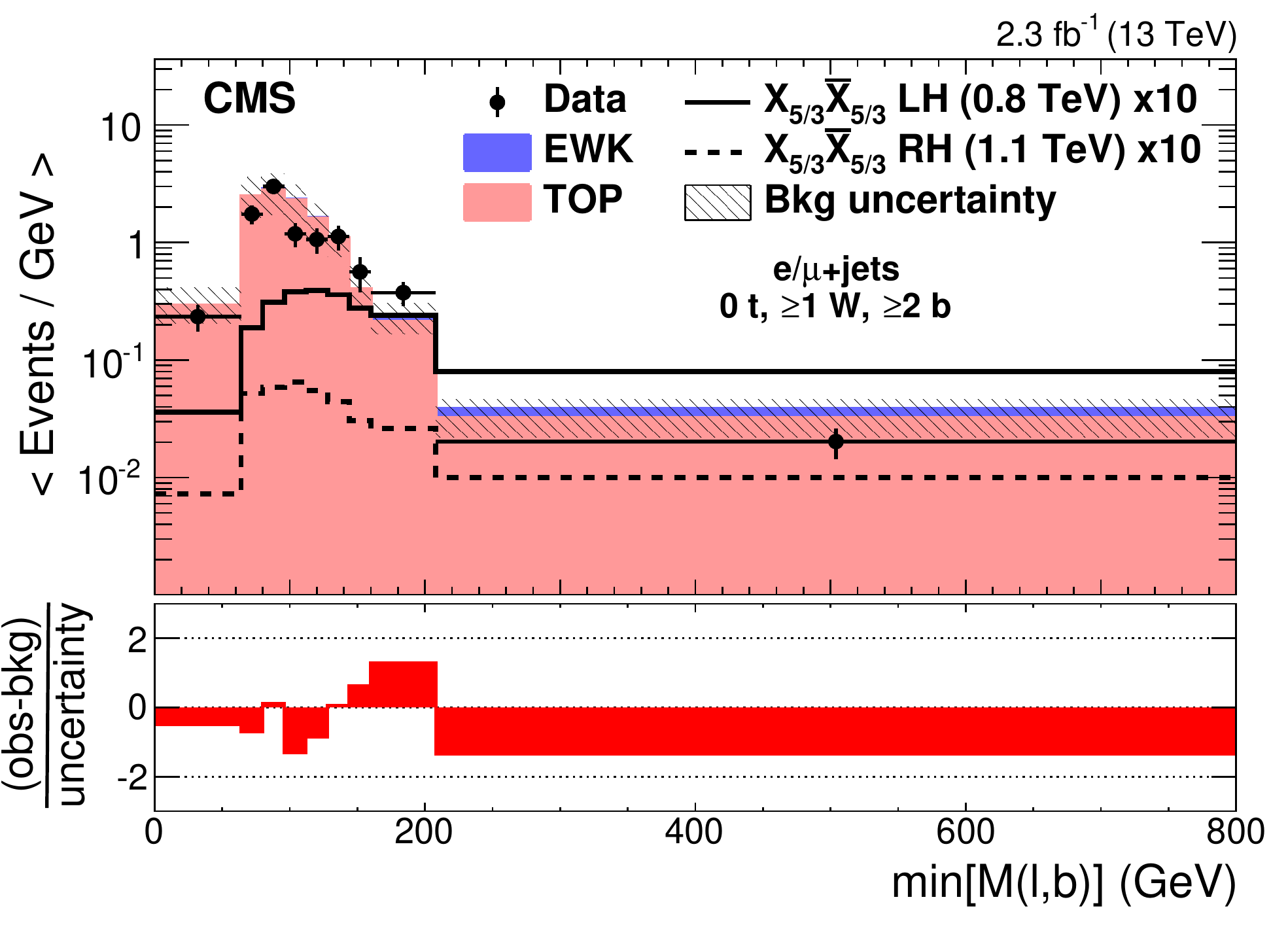}
\caption{Distributions of min[M($\ell$,\,$\PQb$)] in (upper) 0 or (lower) $\ge$1 $\PW$-tagged jets and (left) 1 or 
(right) $\ge$2 $\PQb$-tagged jets categories with 0 $\PQt$-tagged jets for combined electron and muon samples, 
at the final selection level.
The horizontal bars on the data points indicate the bin widths.
The lower panel in all plots shows the difference between the observed and the predicted numbers of events divided by the total uncertainty. 
The total uncertainty is calculated as the sum in quadrature of the statistical uncertainty in the observed measurement and the uncertainty in the background, including both statistical and systematic components.
Also shown are the distributions of representative signal events, which are scaled by a factor of 10.}
\label{fig:Mlb0tcats}
\end{center}
\end{figure}

\begin{figure}[hbtp]
\begin{center}
\includegraphics[width=0.49\textwidth]{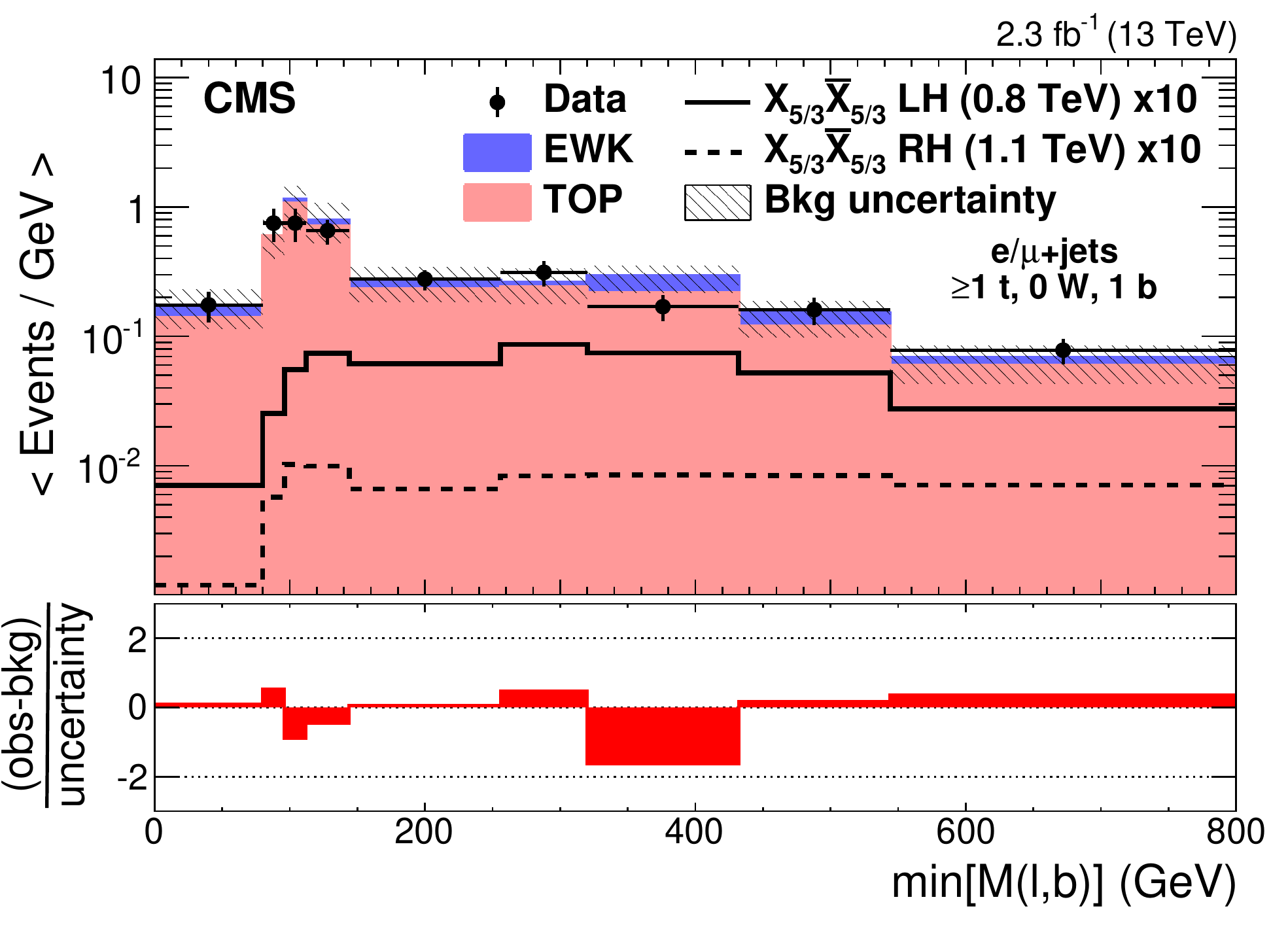}
\includegraphics[width=0.49\textwidth]{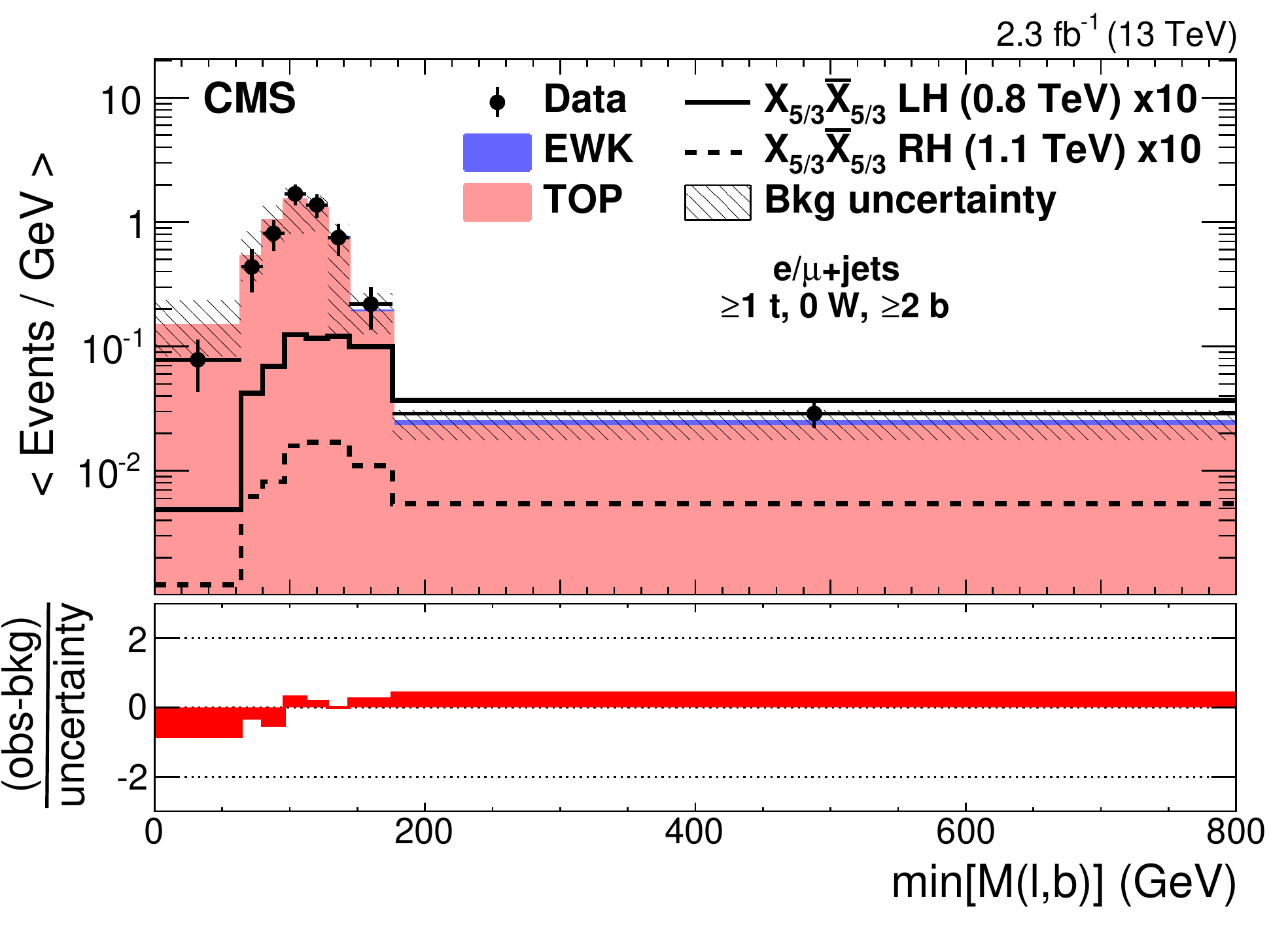}
\includegraphics[width=0.49\textwidth]{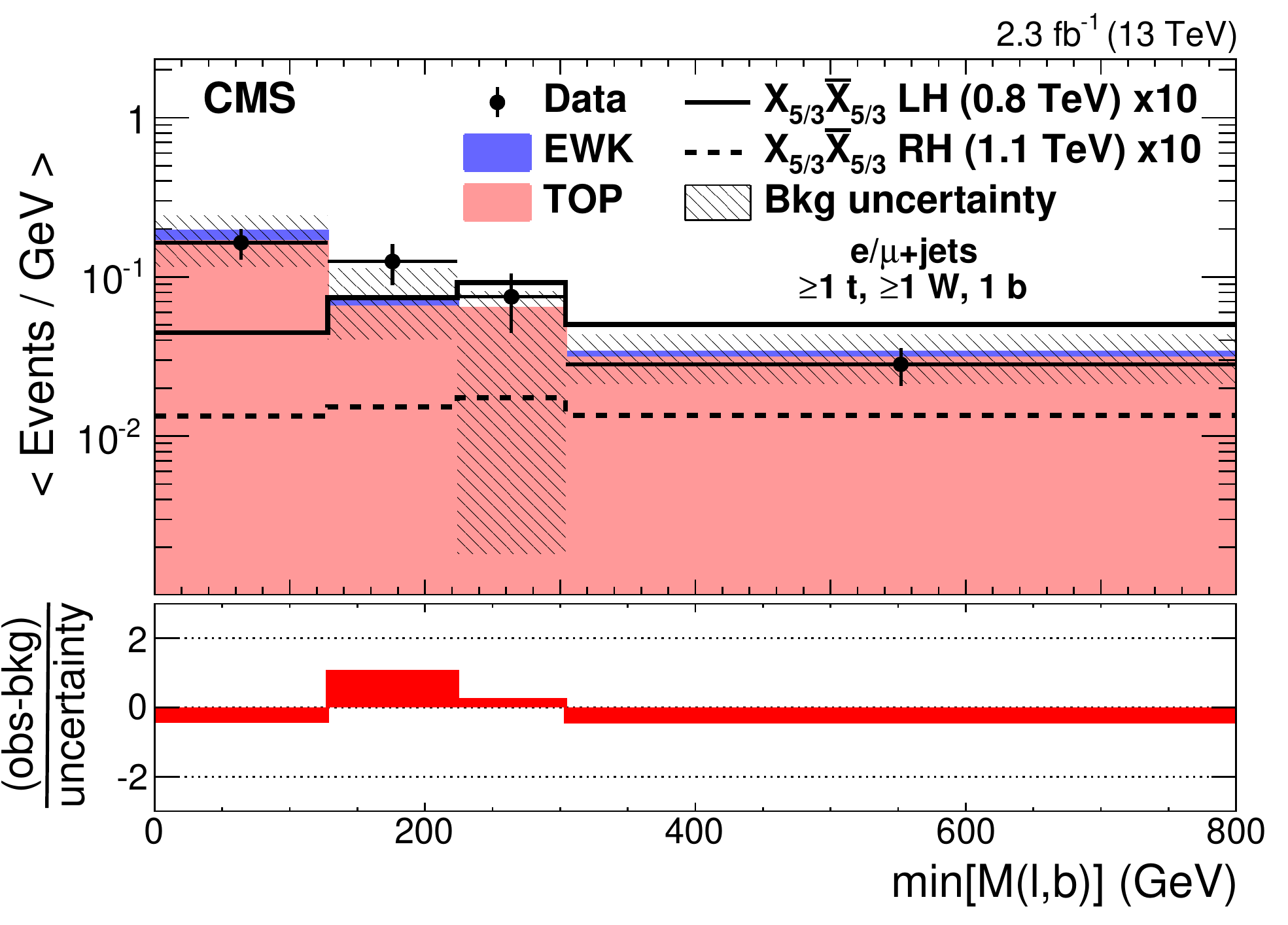}
\includegraphics[width=0.49\textwidth]{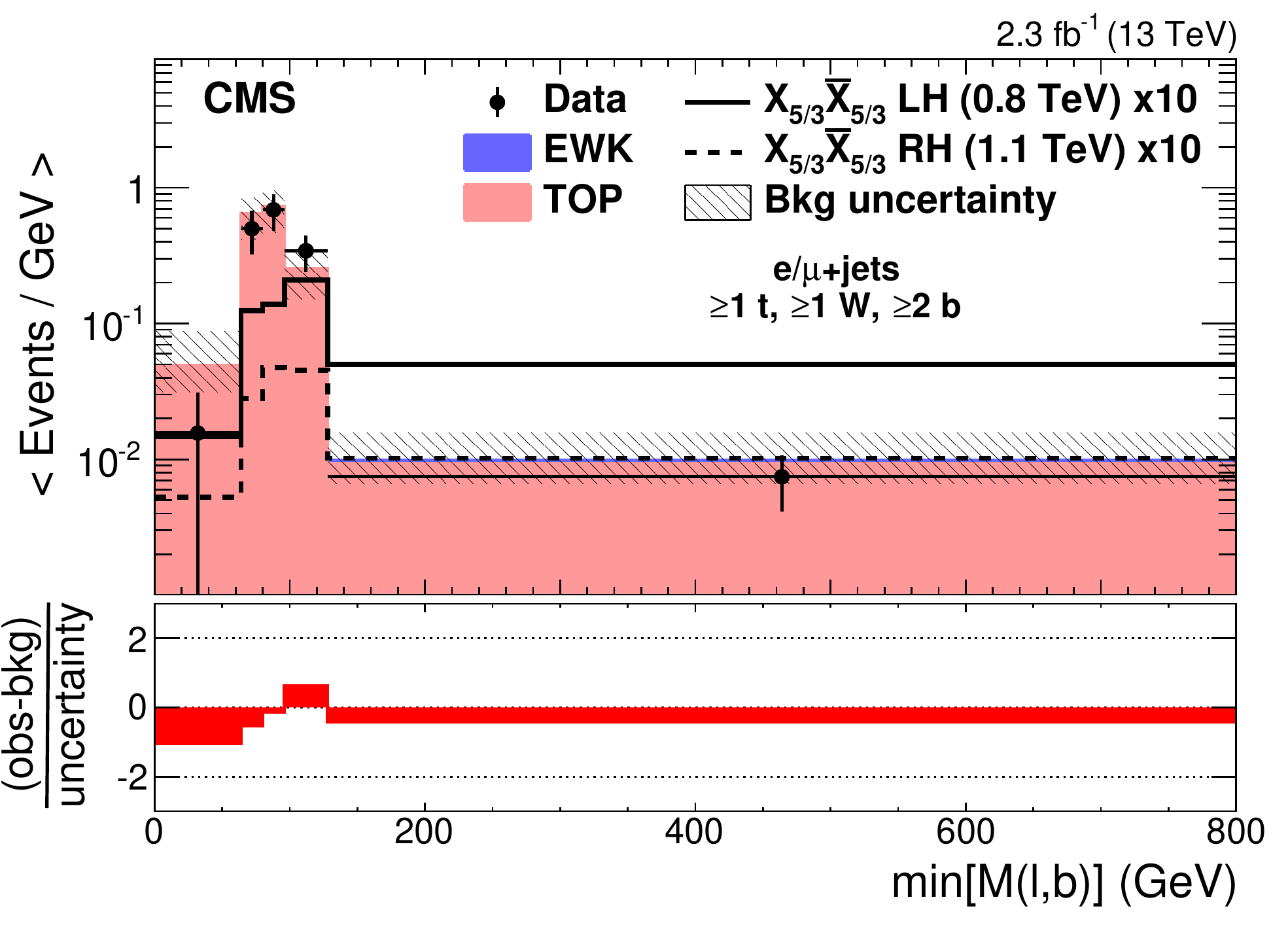}
\caption{Distributions of min[M($\ell$,\,$\PQb$)] in  0 (upper) and $\ge$1 (lower) $\PW$-tagged jets and 1 (left) and 
$\ge$2 (right) $\PQb$-tagged jets categories with $\ge$1 $\PQt$-tagged jets for combined electron and muon samples, 
at the final selection level.
The horizontal bars on the data points indicate the bin widths.
The lower panel in all plots shows the difference between the observed and the predicted numbers of events divided by the total uncertainty. 
The total uncertainty is calculated as the sum in quadrature of the statistical uncertainty in the observed measurement and the uncertainty in the background, including both statistical and systematic components.
Also shown are the distributions of representative signal events, which are scaled by a factor of 10.}
\label{fig:Mlb1ptcats}
\end{center}
\end{figure}

In order to maximize sensitivity to the presence of a \xft signal, in the single-lepton final state analysis events are divided into 16 categories based on lepton flavor 
($\Pe$, $\mu$), and the numbers of $\PQt$-tagged (0, $\ge$1), $\PW$-tagged (0, $\ge$1), and $\PQb$-tagged (1, $\ge$2) jets.
Event yields after the final selection are given in Table~\ref{tab:neventsfinal}. In Figs.~\ref{fig:Mlb0tcats} 
and~\ref{fig:Mlb1ptcats} we show the distributions of min[M($\ell$,\,$\PQb$)] after the final selections for 
events in eight different event categories, depending on the numbers of $\PQt$-, $\PW$-, and $\PQb$-tagged jets, 
after combining the electron and muon channels.
The observed distributions are well reproduced by the SM predictions in all analysis categories.

\begin{table}
\centering
\topcaption{Expected (observed) numbers of background (data) events passing the final selection requirements, 
in the eight tagging categories after combining electron and muon categories, for the single-lepton channel, with an integrated luminosity of 2.3\fbinv. 
Also shown are the numbers of expected events for an LH \xft with a mass of 800\GeV and an RH \xft with a mass of 1.1\TeV.
Uncertainties quoted in the table include both statistical as well as the systematic components listed in Table~\ref{tab:sys-error}. 
The Poisson uncertainty upper bound (${<}1.8$) is used for the categories where the QCD multijet event yield is zero.}
\begin{tabular}{c|cccc}
\hline
Sample & 0 $\PQt$, 0 $\PW$, 1 $\PQb$ &  0 $\PQt$, 0 $\PW$, $\ge$2 $\PQb$ &  0 $\PQt$, $\ge$1 $\PW$, 1 $\PQb$ &  0 $\PQt$, $\ge$1 $\PW$, $\ge$2 $\PQb$ \\ 
\hline
LH \xft (0.8\TeV)    & 3.75 $\pm$ 0.31           & 3.35 $\pm$ 0.35           & 10.75 $\pm$ 0.58\x          & 9.16 $\pm$ 0.72        \\                
RH \xft (1.1\TeV)    & 0.453 $\pm$ 0.043         & 0.329 $\pm$ 0.039         & 1.71 $\pm$ 0.10           & 1.25 $\pm$ 0.11         \\                
\hline																	
TOP                   & 490 $\pm$ 140             & 300 $\pm$ 80\x              & 342 $\pm$ 98\x              & 219 $\pm$ 64\x          \\                
EWK                   & 132 $\pm$ 29\x              & 15.4 $\pm$ 5.7\x            & 53 $\pm$ 14               & 6.6 $\pm$ 3.6         \\                
QCD                   & 2.1 $\pm$ 2.0             & ${<}1.8$                    & ${<}1.8$                    & ${<}1.8$   \\                
\hline
Total bkg.            & 630 $\pm$ 140             & 316 $\pm$ 84\x              & 395 $\pm$ 99\x              & 226 $\pm$ 64\x          \\                
\hline
Data                  & 644                       & 290                       & 366                       & 184                      \\                
\hline 
\multicolumn{5}{c}{}\\ [-1.5ex]
\hline 
Sample & $\ge$1 $\PQt$, 0 $\PW$, 1 $\PQb$ &  $\ge$1 $\PQt$, 0 $\PW$, $\ge$2 $\PQb$ &  $\ge$1 $\PQt$, $\ge$1 $\PW$, 1 $\PQb$ &  $\ge$1 $\PQt$, $\ge$1 $\PW$, $\ge$2 $\PQb$ \\ 
\hline
LH \xft (0.8\TeV)    & 3.79 $\pm$ 0.28           & 3.41 $\pm$ 0.33           & 4.51 $\pm$ 0.33           & 4.55 $\pm$ 0.41        \\                
RH \xft (1.1\TeV)    & 0.565 $\pm$ 0.046         & 0.486 $\pm$ 0.047         & 1.128 $\pm$ 0.087         & 0.98 $\pm$ 0.10        \\                
\hline																	
TOP                   & 155 $\pm$ 44              & 110 $\pm$ 32              & 48 $\pm$ 15               & 40 $\pm$ 10          \\                
EWK                   & 26.0 $\pm$ 8.1            & 2.3 $\pm$ 1.6             & 5.4 $\pm$ 2.9             & 0.31 $\pm$ 0.31       \\                
QCD                   & 0.057 $\pm$ 0.11          & ${<}1.8$                    & ${<}1.8$                    & ${<}1.8$                \\                
\hline
Total bkg.            & 181 $\pm$ 45              & 113 $\pm$ 32              & 53 $\pm$ 16               & 40 $\pm$ 10          \\                
\hline
Data                  & 167                       & 111                       & 53                        & 36                       \\                
\hline 
\end{tabular} 
\label{tab:neventsfinal}
\end{table}

\section{Systematic uncertainties}
\label{sec:Systematics}
The principal systematic uncertainties that are common to both analyses are presented in this section,
while the uncertainties specific to each analysis are presented in Sections~\ref{sec:ssdl_systs} and~\ref{sec:ljets_systs}. 
The uncertainties in the object selection are derived from uncertainties on the efficiency of the trigger, 
lepton reconstruction, lepton identification and isolation. 
These uncertainties are derived from the tag-and-probe studies mentioned in Section~\ref{sec:Reconstruction} 
and are summarized in Table~\ref{tab:LepSys}. Lepton identification and isolation 
uncertainties are applied per lepton, while trigger uncertainties are applied per event.
We also include a 2.3\% uncertainty in the luminosity measurement~\cite{CMS-PAS-LUM-15-001}.
The above uncertainties are applied only to simulation.

\begin{table}[!htbp]
\centering
\topcaption{Details of systematic uncertainties applied for lepton triggering, identification (``ID''), isolation (``ISO''), and integrated luminosity.}
\label{tab:LepSys}
\begin{tabular}{lcl}
\hline
Source & Value & Application\\
\hline
Electron ID & 1\% & per electron\\
Electron ISO & 1\% & per electron\\
Electron trigger & 5\% & per event\\
Electron-electron trigger & 3\% & per event\\
Muon ID & 1\% & per muon\\
Muon ISO & 1\% & per muon\\
Muon trigger & 5\% & per event\\
Muon-muon trigger & 3\% & per event\\
Electron-muon trigger & 3\% & per event\\
Integrated luminosity  &  2.3\%  & per event\\
\hline
\end{tabular}
\end{table}

The uncertainties that can affect the shape of the distributions,
in particular those related to the jet energy scale (JES) and the jet energy resolution (JER), are assessed 
by varying the relevant parameters up and down by one standard deviation (s.d.) and repeating the analysis.
The PDF uncertainty is evaluated using the complete set of NNPDF 3.0 PDF eigenvectors,
following the prescription described in Ref.~\cite{Butterworth:2015oua}.
The uncertainty due to the renormalization and factorization scales
is taken into account by varying the scales up or down by a factor of two and taking the maximum
variation. The uncertainty due to the pileup distribution in the simulation 
is assessed by varying the total inelastic cross section
used in the pileup reweighting by ${\pm}5\%$.

The theoretical uncertainties due to the factorization and renormalization scales and the PDFs 
lead to negligible uncertainties in the signal acceptance in the same-sign dilepton channel.
The single-lepton channel considers the shape variations
in the signal distributions as a result of these uncertainties.

\subsection{The same-sign dilepton final state}
\label{sec:ssdl_systs}
The uncertainties for simulated events are summarized in Table~\ref{tab:MCUncert}, which 
includes uncertainties related to jet energy scale, jet energy resolution, pileup, and the overall normalization uncertainty for each simulated background sample. 
The normalization uncertainty takes into account the uncertainty in the cross section and the uncertainty related 
to the PDFs used to generate the samples. For the rare backgrounds that have either not been observed, or not well measured, 
we assume a conservative normalization uncertainty of 50\%. 
We see variations of up to 2\% for JER and up to 6\% for pileup for some of the simulated background samples. 
For the signal, the JES, JER, and pileup uncertainties in the acceptance correspond to 5\%, 3\%, and 1\%, respectively.

\begin{table}[!htbp]
\centering
\topcaption{Systematic uncertainties in the same-sign dilepton final state, associated with the simulated processes. The ``Normalization'' column refers to uncertainties from the cross section normalization and the choice of PDF.}
\label{tab:MCUncert}
\begin{tabular}{c|cccc}
\hline
Process & JES &JER & Pileup &Normalization \\
\hline
$\ttbar\PW$ & 2\% & 2\% & 6\% & 18\%\\
$\ttbar\PZ $& 3\% & 2\% & 6\% &11\%\\
$\ttbar\PH$ & 4\% & 2\% & 6\% &12\%\\
$\ttbar\ttbar$& 2\%&2\% & 6\% &50\%\\
$\PW\PZ$ & 10\% & 2\% & 6\% &12\%\\
$\PZ\PZ$ & 7\% & 2\% & 6\% &12\%\\
$\PW\PW$ & 6\%& 2\% & 6\% & 50\% \\
$\PW\PW\PZ$ & 7\% & 2\% & 6\% &50\%\\
$\PW\PZ\PZ$ & 9\% & 2\% & 6\% &50\%\\
$\PZ\PZ\PZ$ & 9\% & 2\% & 6\% &50\%\\
\xft & 5\% & 3\% & 1\% & \text{---} \\
\hline
\end{tabular}
\end{table}

As described in Sections~\ref{sec:chargemisid} and~\ref{sec:fakebkg}, 
we also include a 30\% uncertainty for the charge misidentification probability and a 
50\% uncertainty associated with the estimation of the NonPrompt background.
The latter is the dominant source of uncertainty in the total background prediction.

\subsection{The single-lepton final state}
\label{sec:ljets_systs}
The sources of uncertainties in the single-lepton final state are classified according to their effect: 
having the potential to modify normalizations only, shapes only, or both normalizations and shapes.
The uncertainties that affects the normalizations only are listed in Table~\ref{tab:LepSys}.

To model the uncertainties that alter shapes, we consider uncertainties related to the JES, JER, 
$\PQb$ tagging and light quark mistagging efficiencies, $\PW$ tagging uncertainties, $\PQt$ tagging uncertainties, 
event pileup conditions, PDFs, and renormalization, factorization, and parton shower energy scales. 
The effect of reweighting the top quark \pt distribution in $\ttbar$ events,
following the prescription of~\cite{TOPPT},
is considered as a one-sided systematic uncertainty.
The $\ttbar$ and single top parton shower energy scale uncertainties are assessed by 
independently varying the scales up and down by a factor of two.
A summary of these systematic uncertainties, and how they are applied to signal 
and background samples is given in Table~\ref{tab:sys-error}.
In the single-lepton channel the uncertainties in the simulated background processes are 
dominated by the renormalization and factorization scale uncertainties.

\begin{table}[h!tb]
\small
\begin{center}
\topcaption{Summary of all systematic uncertainties considered in the single-lepton channel.
Each uncertainty is included in both signal and all background processes unless noted otherwise.}
\begin{tabular}{lcc}
\hline
Source            &  Uncertainty  & Comment\\
\hline\hline
Shape and normalization & & \\ 
\hline
JES    & ${\pm}1$ s.d. $(\pt,\eta)$  & \\
JER    & ${\pm}1$ s.d. $(\eta)$  & \\
b/c tagging    & ${\pm}1$ s.d. $(\pt)$  & \\
Light quark mistagging    & ${\pm}1$ s.d.  & \\
W tagging: mass resolution & ${\pm}1$ s.d. $(\eta)$ & \\
W tagging: mass scale & ${\pm}1$ s.d. $(\pt,\eta)$ & \\
W tagging: $\tau_2/\tau_1$ & ${\pm}1$ s.d. & \\
t tagging  & ${\pm}1$ s.d. & \\
Pileup    & $\sigma_{\mathrm{inel.}}\pm5$\%   & \\
PDF    & ${\pm}1$ s.d.   & Only for background\\
Renorm./fact. energy scale  & Envelope (${\times}2$, ${\times}0.5$)   & Only for background\\
Parton shower scale & Envelope (${\times}2$, ${\times}0.5$) & Only for $\ttbar$ and single top \\
Top quark \pt & $\Delta$ (weighted, nominal) & Only for $\ttbar$\\
\hline
Shape only & & \\ 
\hline
PDF    & ${\pm}1$ s.d.   & Only for signal\\
Renorm./fact. energy scale  & Envelope (${\times}2$, ${\times}0.5$)   & Only for signal\\
\hline
\end{tabular}
\label{tab:sys-error}
\end{center}
\end{table}

\section{Results}

\begin{figure}[!htbp]
\centering
\includegraphics[width=0.49\textwidth]{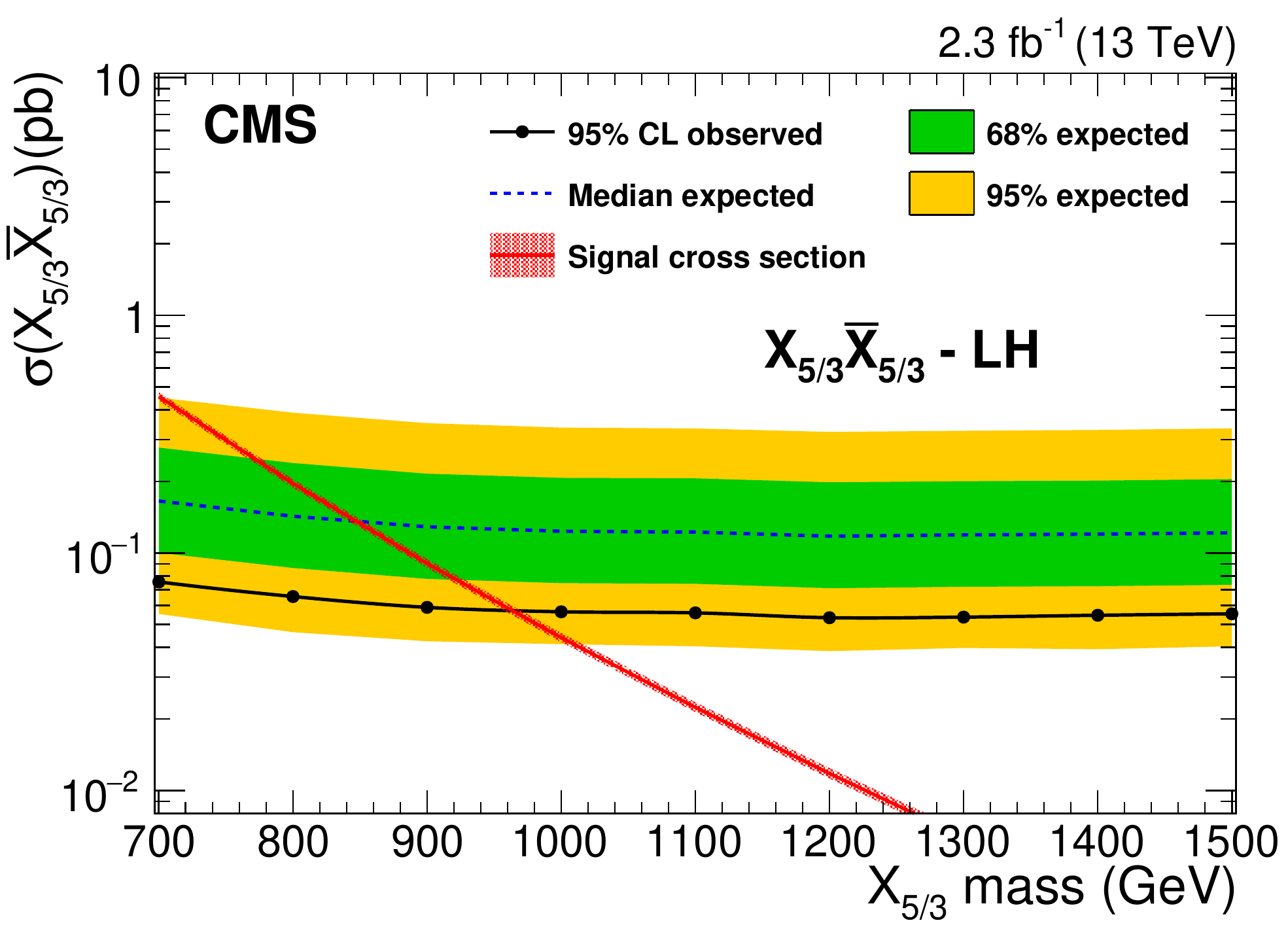}
\includegraphics[width=0.49\textwidth]{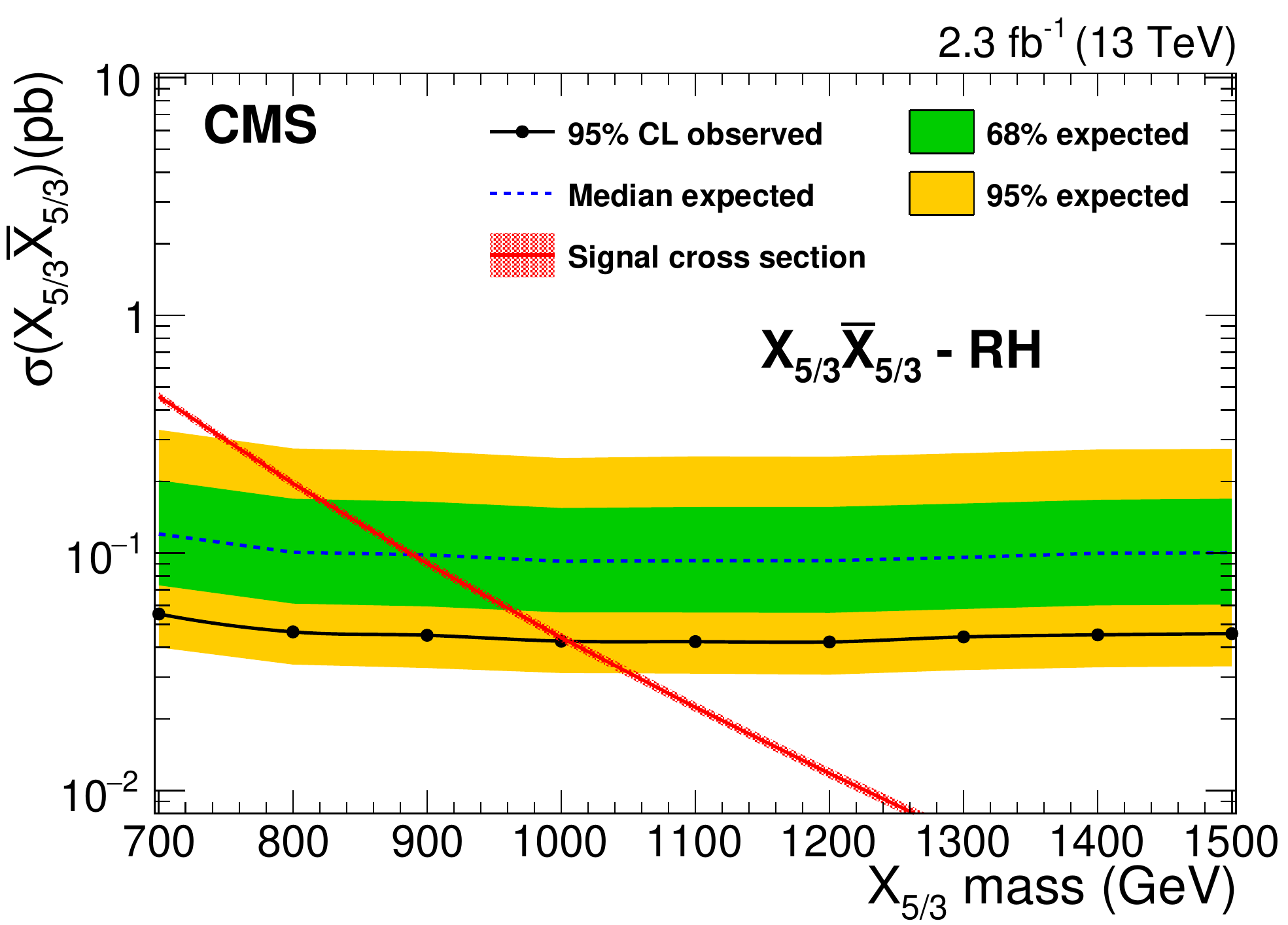} \\
\includegraphics[width=0.49\textwidth]{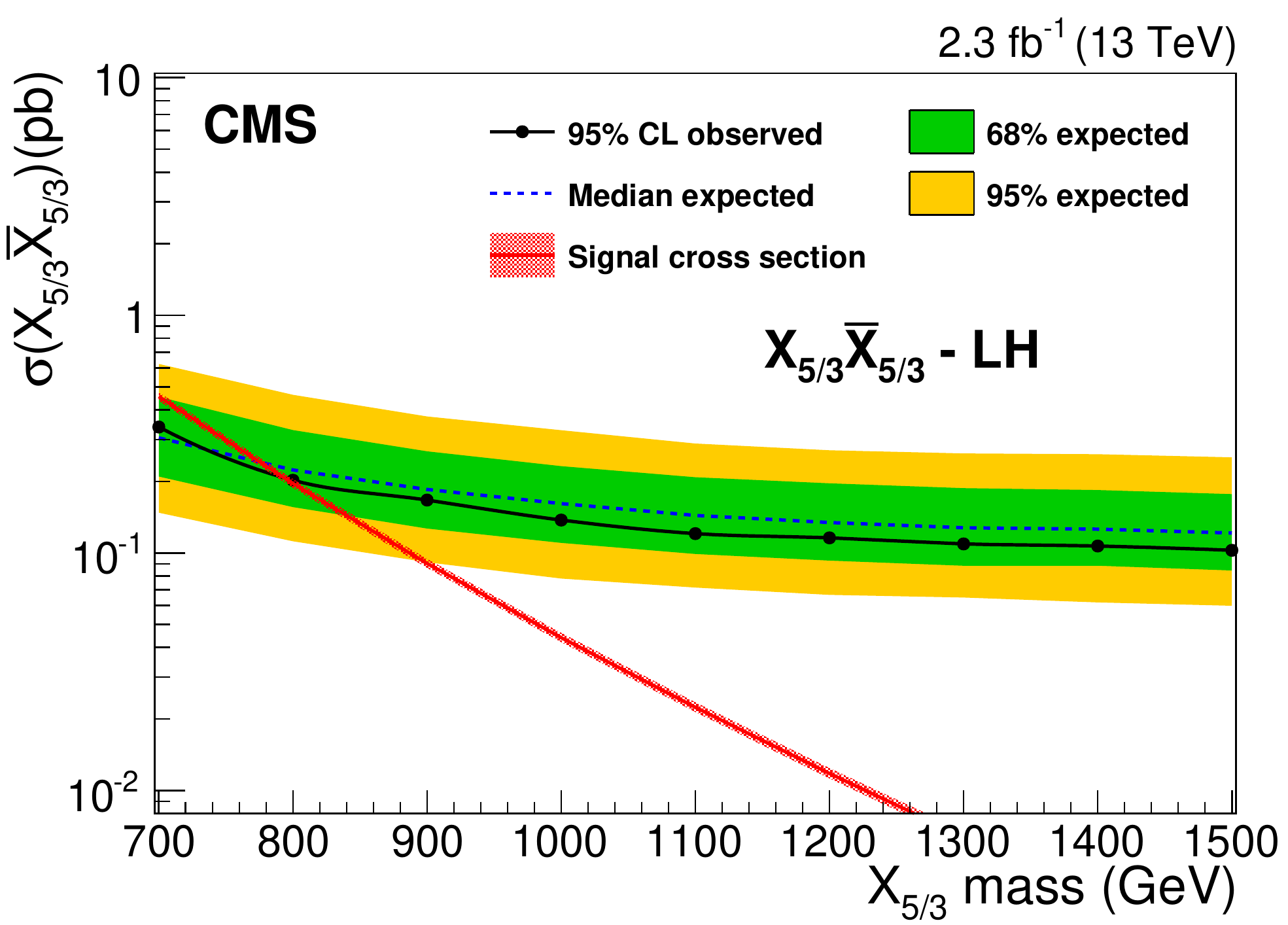}
\includegraphics[width=0.49\textwidth]{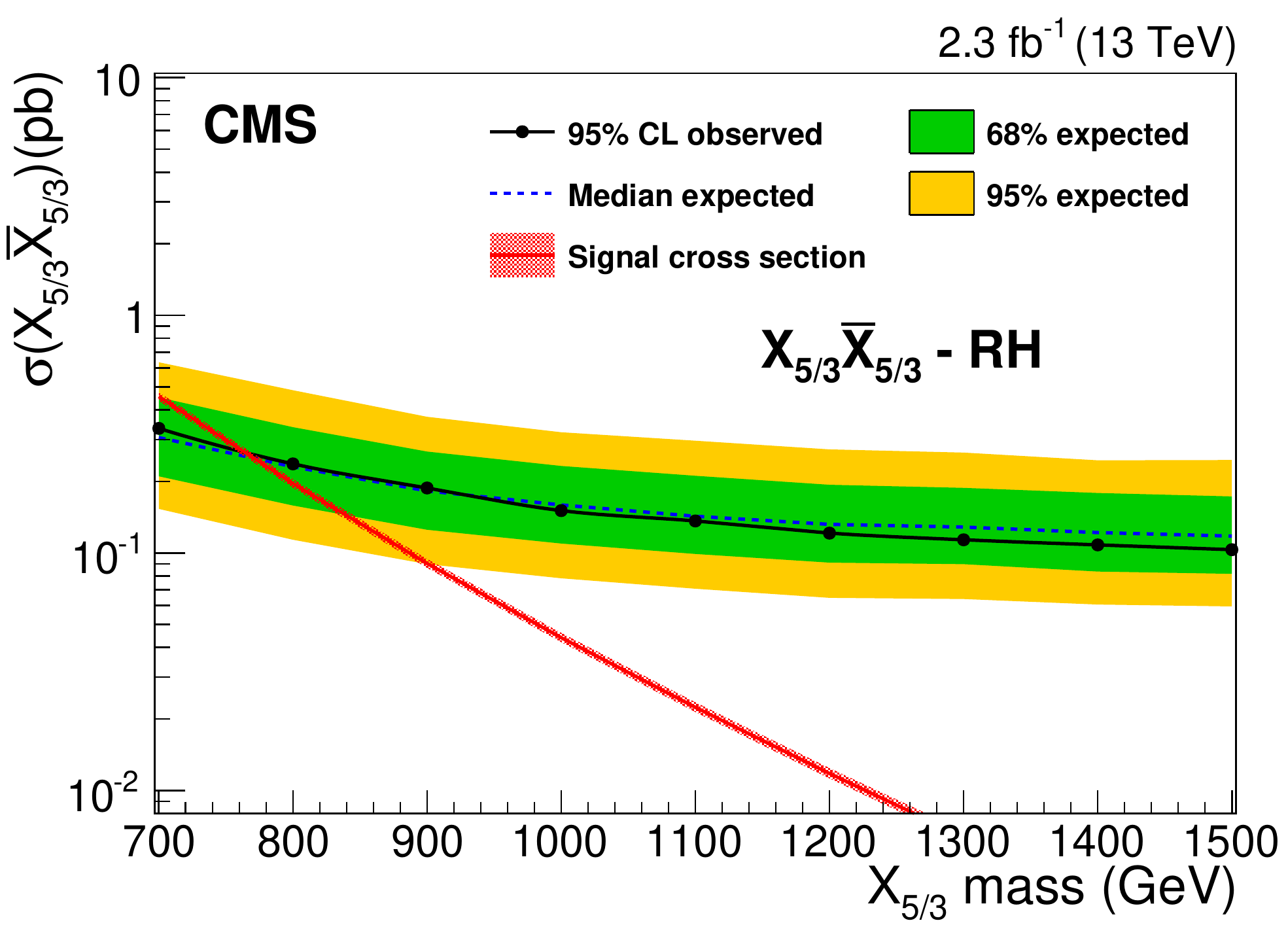}
\caption{The expected and observed upper limits at 95\% CL for a left-handed (left)  and right-handed (right) \xft for the same-sign dilepton signature (upper) and the single-lepton signature (lower) after combining all channels in each signature.
The theoretical prediction for the \xft pair production cross section is shown as a band including its uncertainty.}
\label{fig:Limits}
\end{figure}

We find no significant excess in the data compared to the SM expectations 
and therefore proceed to set 95\% CL upper limits on the production cross section for 
$ \Pp\Pp \rightarrow \mathrm{X}_{5/3}\overline{\mathrm{X}}_{5/3} \rightarrow \PQt\PWp\PAQt\PWm$.
Expected and observed limits are calculated using Bayesian statistics~\cite{THETA} 
with a flat prior distribution in the signal cross section, for both LH and RH \xft scenarios. 
The same-sign dilepton analysis uses a counting experiment to derive limits based on the full set of requirements detailed above, while the single-lepton channel uses a binned likelihood fit to the distribution of the min[M($\ell,\,\PQb$)] variable.
Systematic uncertainties are represented as nuisance parameters with log-normal priors for normalization uncertainties,
Gaussian priors for shape uncertainties with results obtained via the maximum-likelihood value on the signal cross section.
Using the full set of analysis selection criteria and an integrated luminosity of 2.3\fbinv, 
we obtain observed (expected) limits of 1000 (890)\GeV for an RH \xft and 970 (860)\GeV 
for an LH \xft at 95\% CL in the same-sign dilepton channel.
Using the single-lepton channel, the observed (expected) limits are found to be 770 (780)\GeV for an 
RH \xft and 800 (780)\GeV for an LH $\mathrm{X}_{5/3}$, again at 95\% CL.
Both the expected and the observed limits after combining all categories in each signature are 
shown in Fig.~\ref{fig:Limits}, where the PDF, and renormalization and factorization scale uncertainties in
the signal cross section are shown as the band around the theoretical predictions.
The observed limit being consistently lower than the expected limit for the same-sign dilepton results in figure~\ref{fig:Limits} is simply due to the analysis requirements being independent of signal mass.

\begin{figure}[hbtp]
\begin{center}
\includegraphics[width=0.49\textwidth]{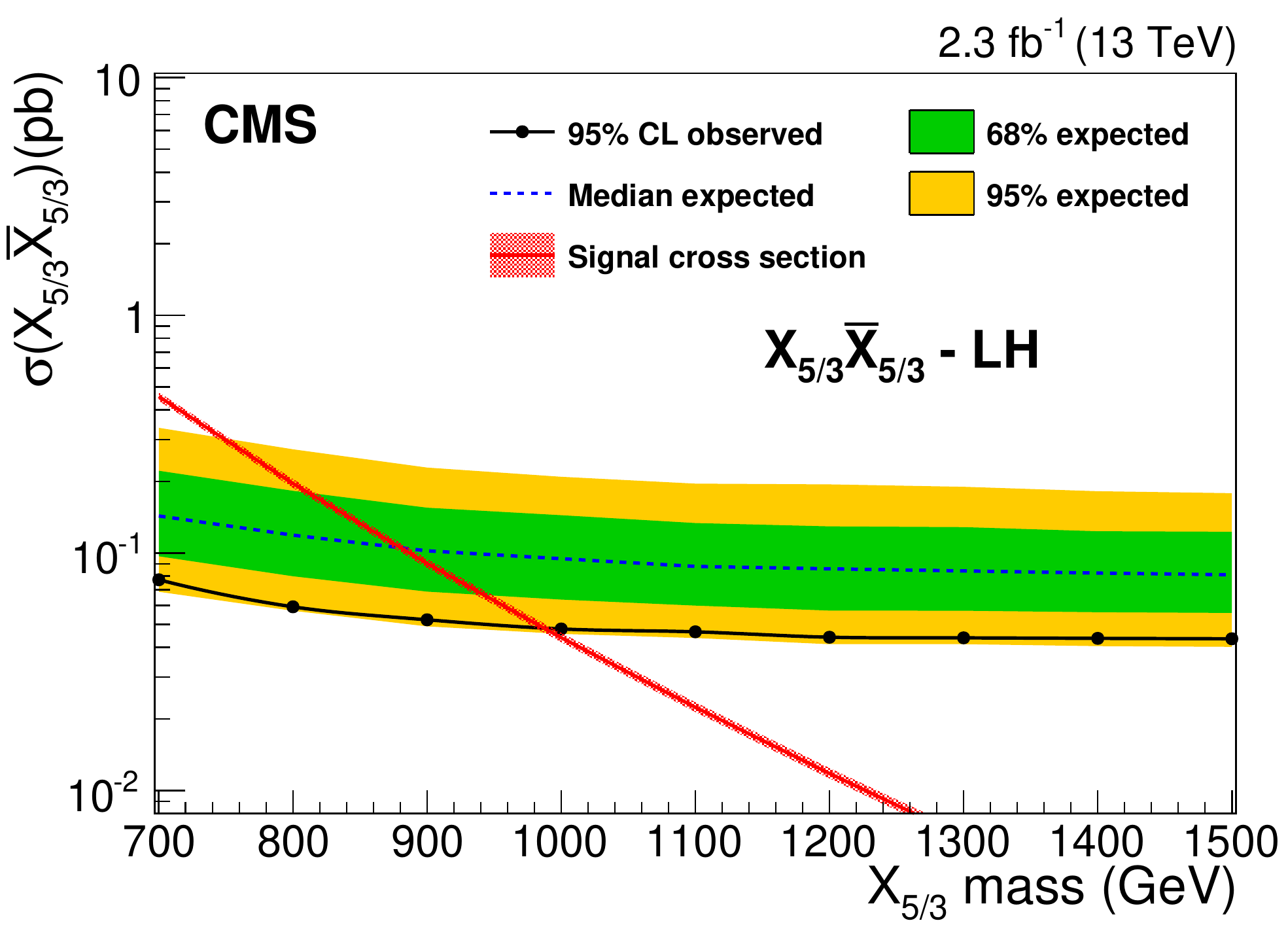}
\includegraphics[width=0.49\textwidth]{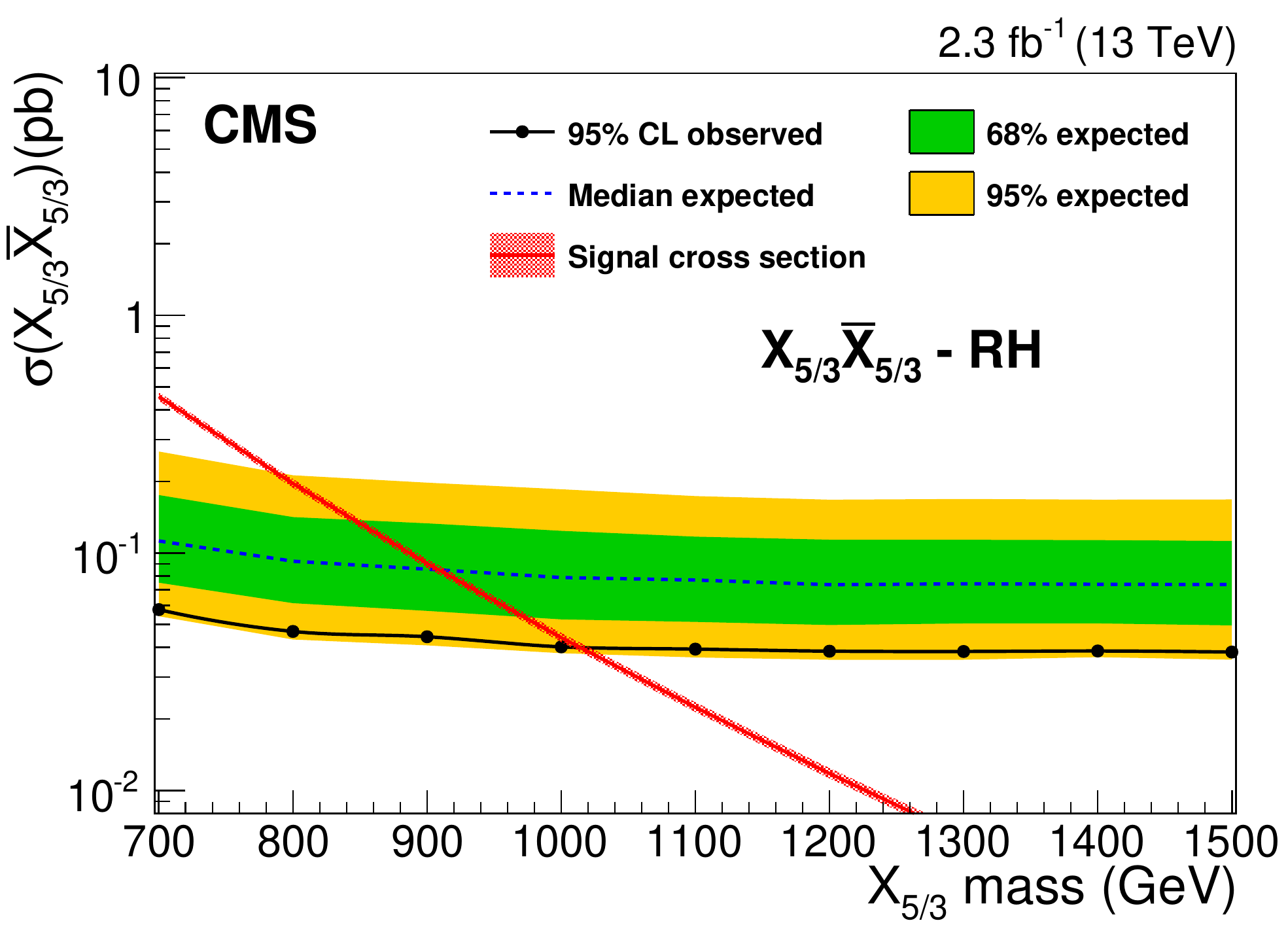}
\caption{The expected and observed upper limits at 95\% CL after combining the same-sign dilepton 
and the single-lepton signatures for left-handed (left) 
and right-handed (right) \xft scenarios. 
The theoretical prediction for the \xft pair production cross section is shown as a band including its uncertainty.}
\label{fig:combo}
\end{center}
\end{figure}

A combination of the results from the analyses of the two final states discussed in this paper,
same-sign dilepton and the single-lepton signatures, is shown in Fig.~\ref{fig:combo}.
In the combination, the observed (expected) exclusion limit on the mass of an RH \xft is found to be 1020 (910)\GeV. 
For the LH \xft signal, the observed (expected) lower limit on the mass is 990 (890)\GeV.

\section{Summary}
A search has been performed for the production of heavy partners of the top quark with charge 5/3
decaying into a top quark and a W boson, 
using 2.3\fbinv of proton-proton collision data collected by the CMS experiment at 13\TeV.
Events with two different signatures are analyzed: 
final states with either a pair of same-sign leptons or a single lepton, along with jets. 
No significant excess is observed in the data above the expected standard model background. 
Upper bounds at 95\% confidence level are set on the production cross section of heavy top quark partners.
The \xft masses with right-handed (left-handed) couplings below 1020 (990)\GeV are excluded at 95\% confidence level.
These are the most stringent limits placed on the \xft mass and the first limits based on a combination of these two different final states.

\begin{acknowledgments}
We congratulate our colleagues in the CERN accelerator departments for the excellent performance of the LHC and thank the technical and administrative staffs at CERN and at other CMS institutes for their contributions to the success of the CMS effort. In addition, we gratefully acknowledge the computing centres and personnel of the Worldwide LHC Computing Grid for delivering so effectively the computing infrastructure essential to our analyses. Finally, we acknowledge the enduring support for the construction and operation of the LHC and the CMS detector provided by the following funding agencies: BMWFW and FWF (Austria); FNRS and FWO (Belgium); CNPq, CAPES, FAPERJ, and FAPESP (Brazil); MES (Bulgaria); CERN; CAS, MoST, and NSFC (China); COLCIENCIAS (Colombia); MSES and CSF (Croatia); RPF (Cyprus); SENESCYT (Ecuador); MoER, ERC IUT, and ERDF (Estonia); Academy of Finland, MEC, and HIP (Finland); CEA and CNRS/IN2P3 (France); BMBF, DFG, and HGF (Germany); GSRT (Greece); OTKA and NIH (Hungary); DAE and DST (India); IPM (Iran); SFI (Ireland); INFN (Italy); MSIP and NRF (Republic of Korea); LAS (Lithuania); MOE and UM (Malaysia); BUAP, CINVESTAV, CONACYT, LNS, SEP, and UASLP-FAI (Mexico); MBIE (New Zealand); PAEC (Pakistan); MSHE and NSC (Poland); FCT (Portugal); JINR (Dubna); MON, RosAtom, RAS, RFBR and RAEP (Russia); MESTD (Serbia); SEIDI, CPAN, PCTI and FEDER (Spain); Swiss Funding Agencies (Switzerland); MST (Taipei); ThEPCenter, IPST, STAR, and NSTDA (Thailand); TUBITAK and TAEK (Turkey); NASU and SFFR (Ukraine); STFC (United Kingdom); DOE and NSF (USA).

\hyphenation{Rachada-pisek} Individuals have received support from the Marie-Curie programme and the European Research Council and Horizon 2020 Grant, contract No. 675440 (European Union); the Leventis Foundation; the A. P. Sloan Foundation; the Alexander von Humboldt Foundation; the Belgian Federal Science Policy Office; the Fonds pour la Formation \`a la Recherche dans l'Industrie et dans l'Agriculture (FRIA-Belgium); the Agentschap voor Innovatie door Wetenschap en Technologie (IWT-Belgium); the Ministry of Education, Youth and Sports (MEYS) of the Czech Republic; the Council of Science and Industrial Research, India; the HOMING PLUS programme of the Foundation for Polish Science, cofinanced from European Union, Regional Development Fund, the Mobility Plus programme of the Ministry of Science and Higher Education, the National Science Center (Poland), contracts Harmonia 2014/14/M/ST2/00428, Opus 2014/13/B/ST2/02543, 2014/15/B/ST2/03998, and 2015/19/B/ST2/02861, Sonata-bis 2012/07/E/ST2/01406; the National Priorities Research Program by Qatar National Research Fund; the Programa Clar\'in-COFUND del Principado de Asturias; the Thalis and Aristeia programmes cofinanced by EU-ESF and the Greek NSRF; the Rachadapisek Sompot Fund for Postdoctoral Fellowship, Chulalongkorn University and the Chulalongkorn Academic into Its 2nd Century Project Advancement Project (Thailand); and the Welch Foundation, contract C-1845.

\end{acknowledgments}

\bibliography{auto_generated}

\appendix

\cleardoublepage \appendix\section{The CMS Collaboration \label{app:collab}}\begin{sloppypar}\hyphenpenalty=5000\widowpenalty=500\clubpenalty=5000\textbf{Yerevan Physics Institute,  Yerevan,  Armenia}\\*[0pt]
A.M.~Sirunyan, A.~Tumasyan
\vskip\cmsinstskip
\textbf{Institut f\"{u}r Hochenergiephysik,  Wien,  Austria}\\*[0pt]
W.~Adam, E.~Asilar, T.~Bergauer, J.~Brandstetter, E.~Brondolin, M.~Dragicevic, J.~Er\"{o}, M.~Flechl, M.~Friedl, R.~Fr\"{u}hwirth\cmsAuthorMark{1}, V.M.~Ghete, C.~Hartl, N.~H\"{o}rmann, J.~Hrubec, M.~Jeitler\cmsAuthorMark{1}, A.~K\"{o}nig, I.~Kr\"{a}tschmer, D.~Liko, T.~Matsushita, I.~Mikulec, D.~Rabady, N.~Rad, B.~Rahbaran, H.~Rohringer, J.~Schieck\cmsAuthorMark{1}, J.~Strauss, W.~Waltenberger, C.-E.~Wulz\cmsAuthorMark{1}
% ...
\vskip\cmsinstskip
1:~~Also at Vienna University of Technology, Vienna, Austria\\
% ...

\end{sloppypar}
\end{document}